\documentclass[times]{aastex62}

\usepackage{acronym}
\newacro{CDM}{cold dark matter}
\newcommand{\CDM}{\ac{CDM}}
\newacro{FDM}{Fuzzy dark matter}
\newcommand{\FDM}{\ac{FDM}}
\newacro{DF}{distribution function}
\newcommand{\DF}{\ac{DF}}
\newacro{BNUU}{Boltzmann--Nordheim--Uehling--Uhlenbeck}
\newcommand{\BNUU}{\ac{BNUU}}
\newacro{FP}{Fokker--Planck}
\newcommand{\FP}{\ac{FP}}
\newacro{BL}{Balescu--Lenard}
\newcommand{\BL}{\ac{BL}}
\newacro{SP}{Schr\"{o}dinger--Poisson}
\newcommand{\SP}{\ac{SP}}

\usepackage{bm}
\usepackage{mathtools}

\newcommand{\rd}{\mathrm{d}}
\newcommand{\rb}{\mathrm{b}}
\newcommand{\re}{\mathrm{e}}
\newcommand{\ri}{\mathrm{i}}
\newcommand{\rs}{\mathrm{s}}

\newcommand{\mb}{m_{\mathrm{b}}}
\newcommand{\mt}{m_{\mathrm{t}}}

\newcommand{\pc}{\,\mathrm{pc}}
\newcommand{\kpc}{\,\mathrm{kpc}}
\newcommand{\eV}{\,\mathrm{eV}}
\newcommand{\kms}{\,\mathrm{km\ s}^{-1}}
\newcommand{\half}{\tfrac{1}{2}}
\newcommand{\Msun}{M_{\odot}}
\newcommand{\p}{\partial}
\newcommand{\oD}{\mathfrak{D}}
\newcommand{\hoD}{\widehat{\oD}}
\newcommand{\erf}{\mathrm{erf}}
\newcommand{\Li}{\mathrm{Li}}
\renewcommand{\bv}{\mathbf{v}}
\newcommand{\bk}{\mathbf{k}}
\newcommand{\br}{\mathbf{r}}
\newcommand{\bs}{\mathbf{s}}
\newcommand{\bp}{\mathbf{p}}
\newcommand{\ba}{\mathbf{a}}
\newcommand{\bb}{\mathbf{b}}
\newcommand{\bq}{\mathbf{q}}
\newcommand{\deltaD}{\delta_{\mathrm{D}}}
\newcommand{\meff}{m_\mathrm{eff}}
\newcommand{\sigmaeff}{\sigma_\mathrm{eff}}
\newcommand{\hPhi}{\widehat{\Phi}}
\newcommand{\hk}{\widehat{k}}
\newcommand{\hbk}{\widehat{\mathbf{k}}}
\newcommand{\hbvp}{\widehat{\mathbf{v}}^{\prime}}
\newcommand{\brp}{\mathbf{r}^{\prime}}

\newcommand{\bkp}{\mathbf{k}^{\prime}}
\newcommand{\omegap}{\omega^{\prime}}
\newcommand{\bvp}{\mathbf{v}^{\prime}}
\newcommand{\bvpp}{\mathbf{v}^{\prime\prime}}
\newcommand{\eps}{\epsilon}
\newcommand{\veps}{\varepsilon}
\newcommand{\vphi}{\varphi}
\newcommand{\Fb}{F_{\rb}}
\newcommand{\Feff}{F_{\mathrm{eff}}}

\newcommand{\lbarsig}{\lambdabar_\sigma}
\newcommand{\hf}{\widehat{f}}
\newcommand{\hC}{\widehat{C}}
\newcommand{\oW}{\overline{W}}
\newcommand{\mP}{\mathcal{P}}
\newcommand{\bnabla}{\bm{\nabla}}
\newcommand{\kc}{k_{\mathrm{c}}}
\newcommand{\kJ}{k_{\mathrm{J}}}
\newcommand{\tkJ}{\tilde{k}_{\mathrm{J}}}
\newcommand{\td}{t_{\mathrm{d}}}
\newcommand{\trelax}{t_{\mathrm{relax}}}
\newcommand{\qmin}{q_{\mathrm{min}}}
\newcommand{\qmax}{q_{\mathrm{max}}}
\newcommand{\vp}{v^{\prime}}
\newcommand{\sigmac}{\sigma_{\mathrm{c}}}
\newcommand{\betac}{\beta_{\mathrm{c}}}
\newcommand{\Dc}{D^{\mathrm{c}}}
\newcommand{\Db}{D^{\mathrm{b}}}
\newcommand{\mB}{\mathcal{B}}

\newcommand{\ImPart}{\mathrm{Im}}
\newcommand{\omegaR}{\omega_{\mathrm{R}}}
\newcommand{\Fd}{F_{\mathrm{d}}}
\newcommand{\kp}{k^{\prime}}
\newcommand{\bsigma}{\bm{\sigma}}
\newcommand{\bdelta}{\bm{\delta}}
\newcommand{\kmin}{k_{\mathrm{min}}}
\newcommand{\kmax}{k_{\mathrm{max}}}

\newcommand{\bn}{\mathbf{n}}
\newcommand{\hpsi}{\widehat{\psi}}
\newcommand{\vmax}{v_{\mathrm{max}}}
\newcommand{\Fpara}{F_{\parallel}}
\newcommand{\cs}{c_{\mathrm{s}}}
\newcommand{\kcs}{k_{c_{\mathrm{s}}}}

\usepackage{soul}

\shorttitle{Fuzzy dark matter}
\shortauthors{Bar-Or, Fouvry and Tremaine}

\begin{document}

\title{Relaxation in a Fuzzy Dark Matter Halo. II.
Self-consistent kinetic equations}

\author[0000-0002-8927-4571]{Ben Bar-Or}
\affiliation{Institute for Advanced Study, Princeton, NJ 08540, USA}

\author[0000-0002-0030-371X]{Jean-Baptiste Fouvry}
\affiliation{Institut d'Astrophysique de Paris, and UPMC Univ. Paris 06, (UMR7095), 98 bis Boulevard Arago, 75014 Paris, France}
\affiliation{Institute for Advanced Study, Princeton, NJ 08540, USA}

\author[0000-0002-0278-7180]{Scott Tremaine}
\affiliation{Institute for Advanced Study, Princeton, NJ 08540, USA}
\affiliation{Canadian Institute for Theoretical Astrophysics, University of Toronto, 60 St.\ George Street,
Toronto, ON M5S 3H8, Canada}

\begin{abstract}
Fuzzy dark matter (FDM) is composed of ultra-light bosons having a de~Broglie wavelength that is comparable to the size of the stellar component of galaxies at typical galactic velocities.
FDM behaves like cold dark matter on large scales.
However, on the scale of the de~Broglie wavelength, an FDM halo exhibits density fluctuations that lead to relaxation, a process similar to the two-body relaxation that occurs in classical gravitational N-body systems and is described by the Fokker--Planck equation.
We derive the FDM analog of that kinetic equation, and solve it to find the evolution of the velocity distribution in a spatially homogeneous FDM halo. We also determine the dielectric function and the dispersion relation for linear waves in an FDM halo.
\end{abstract}

\bigskip
\section{Introduction} 
\label{sec:intro}

Cosmological models based on \CDM\ explain most features of the cosmic microwave background, large-scale structure, and other cosmological phenomena. However, \CDM\ has been less successful in predicting the properties of small-scale structure, such as the abundance of dwarf galaxies and the dark-matter density profiles near the centers of galaxies \citep[e.g.,][]{Weinberg+2015,Bullock+2017}. This shortcoming may reflect either our limited understanding of baryonic physics on these scales or deviations of the behavior of the dark matter from the predictions of \CDM\@.
 
\FDM\ is dark matter composed of bosons with mass ${ \mb \simeq 10^{-21} \text{--} 10^{-22} \eV}$, so small that the de~Broglie wavelength
\begin{equation}
   \lambda = \frac{h}{\mb v} = 1.20\kpc \,\frac{10^{-22}\eV}{\mb}\frac{100\kms}{v}
\end{equation}
is comparable to galaxy scales at a typical galaxy velocity $v$ \citep[see, e.g.,][]{Hu+2000, Marsh2016,Hui+2017}. \FDM\ behaves like \CDM\ on scales much larger than the de~Broglie wavelength and thus preserves the success of \CDM\ in explaining the properties of large-scale structure and the cosmic microwave background. However, on small scales \FDM\ behaves quite differently from \CDM\@. In particular, \FDM\ exhibits density fluctuations on the scale of the de~Broglie wavelength that arise from interference patterns. These fluctuations never damp, in contrast to the fluctuations in the density of \CDM\ that arise from incomplete phase mixing. The gravitational field from these fluctuations scatters both condensed baryonic objects---stars, globular clusters, black holes, etc.---and the \FDM\ waves themselves.

\cite{Hui+2017} argued, and~\cite{Bar-Or+2019} (hereafter Paper I) showed explicitly, that scattering of condensed objects by the \FDM\ fluctuations can be analyzed by treating the \FDM\ fluctuations as quasiparticles with effective mass of order the mass contained within the typical angular de~Broglie wavelength, ${ \lbarsig=\hbar/(\mb \sigma) }$ with $\sigma$ the one-dimensional velocity dispersion in the galaxy (see also \citealt{ElZant+2020}). In particular, Paper I computed the diffusion coefficients originating from a homogeneous \FDM\ background that can be used in the classical \FP\ equation to describe the evolution of the \DF\ of a population of stars or other condensed objects.

The goal of this paper is to describe the evolution of the \FDM\ distribution function itself due to scattering by these same fluctuations, by deriving the appropriate wave version of the \FP\ equation. We shall focus on systems that are homogeneous on large scales after averaging over the fluctuations, although our results can be applied to inhomogeneous systems so long as the system size is large compared to the de~Broglie wavelength. Relaxation of the \FDM\ distribution may lead to the formation of a central soliton or Bose--Einstein condensate,
but we do not study the formation of the condensate here. 

The relaxation time of a test particle orbiting in a stellar system of density $\rho_{0}$ and velocity dispersion $\sigma$, composed
of classical particles of mass $m$, is~\citep[][eq.\ 7.106]{Binney+2008}
\begin{equation}
  \label{eq:intro_I}
  \trelax \simeq 0.34\,\frac{\sigma^3}{G^2m \rho_{0} \ln \Lambda},
\end{equation}
where ${ \ln\Lambda }$ is the Coulomb logarithm, with ${ \Lambda \simeq R/b }$ where $R$ is the size of the system and $b$ is the larger of ${ Gm/\sigma^2 }$ and the size of the test particle. In an \FDM\ halo the effective mass is
${ \meff \simeq \rho_{0} \lbarsig^3 }$
so the relaxation of the \FDM\ halo takes place on a timescale 
\begin{equation}
  \label{eq:intro_II}
  \trelax \simeq \frac{\mb^3\sigma^6}{G^2 \rho_{0}^{2} \hbar^3 \ln (R/\lbarsig)}.
\end{equation}
Our aim is to place this approximate result on a solid quantitative foundation. 

Many of our results have appeared already in the literature in several contexts: weak turbulence, the nonlinear Schr\"{o}dinger equation, quantum plasmas, etc. \citep[see, e.g.\@,][]{Levkov+2018}. Nevertheless we have found it simpler and more transparent to provide self-contained derivations.

The present paper is organized as follows.
In Section~\ref{sec:KineticEquations}, we discuss the relaxation of particles and waves. In particular, we present a closed kinetic equation describing the self-consistent relaxation of a homogeneous \FDM\ halo under the effects of its self-generated fluctuations. In Section~\ref{sec:DerivationBNUU}, we present a first heuristic derivation of that kinetic equation
relying on the \BNUU\ equation. In Section~\ref{sec:DerivationKlim}, we revisit that same derivation, this time starting from a quasi-linear expansion of the coupled
\SP\ system. In Section~\ref{sec:Stability}, we discuss the linear stability of \FDM\ halo, and in Section~\ref{sec:Application} we present some applications of this generalized kinetic equation. Finally, we conclude in Section~\ref{sec:Conclusion}.

\bigskip
\section{Relaxation and kinetic equations}
\label{sec:KineticEquations}

We shall work with an infinite halo that is homogeneous in a time-averaged sense, and characterized by a mean density $\rho_0$.  We also invoke the Jeans swindle, that is, we ignore any acceleration due to $\rho_0$ \citep[][\S5.2.2]{Binney+2008},
and focus only on the effects due to the density perturbations.
We define the halo \DF\ $\Fb(\bv)$ such that ${ \Fb(\bv) \rd \br \rd \bv }$ is the time-averaged mass of halo particles in the phase-space volume element ${ \rd \br \rd \bv }$.
Thus the mean density and one-dimensional velocity dispersion are given by 
\begin{equation}
    \rho_0= \!\!\int\!\! \rd \bv \, \Fb(\bv);
    \quad
    3 \rho_0 \sigma^2 = \!\!\int\!\! \rd \bv \, v^2 \Fb(\bv) .
    \label{def_rho}
\end{equation}

In order to highlight the connections between the classical
and quantum cases, we will successively consider the case of the relaxation of classical particles
induced by a classical halo (Section~\ref{sec:PartWithPart}),
then the relaxation of classical particles induced by a fuzzy halo (Section~\ref{sec:PartWithWave}),
and finally the relaxation of a fuzzy halo induced by itself (Section~\ref{sec:WaveWithWave}).

\subsection{Relaxation of particles by particles}
\label{sec:PartWithPart}

First we review relaxation in a halo composed of classical particles of mass $m$.  Let ${ F(\bv) }$ be the \DF\ of a population of point-like classical test objects of mass $\mt$. The evolution of $F(\bv)$ due to interactions between the test objects and the background halo particles is described by the classical Landau equation~\citep{Landau1936, Lifshitz1981,Chavanis2013}, which reads
\begin{equation}
  \label{eq:landau_c}
  \frac{\p F(\bv)}{\p t} = 2 G^2 \ln\Lambda \frac{\p }{\p v_i} \!\int\!\! \rd \bvp \, u_{ij}(\bv - \bvp) 
  \bigg[ m \Fb(\bvp) \frac{\p F(\bv)}{\p v_j}
  - \mt \, \frac{\p \Fb (\bvp)}{\p \vp_{j}} \, F (\bv)
  \bigg] ,
\end{equation}
where
\begin{equation}
  \label{eq:uij_I}
  u_{ij}(\bv) = \!\!\int\!\! \rd \hbk \, \hk_i \hk_j \, \deltaD
  (\hbk \!\cdot\! \bv )
  = \pi\frac{\p^2v}{\p v_i\p v_j}=\pi \frac{v^2 \delta_{ij} - v_i v_j}{v^3},
\end{equation} 
is the collision kernel. Here ${ \hbk = \bk/k }$ is a unit vector along $\bk$
and summation over repeated Cartesian indices is assumed.
The Coulomb logarithm reads ${ \ln\Lambda=\ln (\kmax/\kmin) }$,
with $\kmax$ and $\kmin$ the maximum and minimum wavenumbers
that contribute to the relaxation.
Typically $\kmin^{-1}$ is approximately the size of the halo or the radius of the orbit of the test object, and ${ \kmax \simeq \sigma^2/[G(\mt+m)] }$ corresponds to the scale associated
with strong deflections in two-body encounters.
The first term in Eq.~\eqref{eq:landau_c} represents diffusion and is independent of the mass of the test objects, while the second term represents dynamical friction (also called friction due to polarization), and is independent of the mass of the halo particles at fixed halo density. This equation is equivalent to the standard \FP\ equation
of~\cite{Chandrasekhar1942}.

\subsection{Relaxation of particles by waves}
\label{sec:PartWithWave}

Let us now assume that the diffusion of the classical test particles is being sourced by a background fuzzy halo
composed of ultra-light particles of individual mass $\mb$,
which must therefore be treated as waves.

As shown in Paper I, the Landau equation~\eqref{eq:landau_c} for the evolution of classical test particles becomes
\begin{equation}
  \label{eq:landau_b}
  \frac{\p F (\bv)}{\p t} = 2 G^2 \ln\Lambda \frac{\p }{\p v_i} \!\int\!\! \rd \bvp \, u_{ij}(\bv - \bvp) 
  \bigg\{
  \bigg[\mb + \frac{h^3}{\mb^3}\Fb(\bvp)\bigg]
  \Fb(\bvp)
  \frac{\p F(\bv)}{\p v_j}
 - \mt \, \frac{\p \Fb (\bvp)}{\p \vp_{j}} F(\bv)
  \bigg\};
\end{equation}
here ${ h=2\pi\hbar}$ and the factor $\kmax$ in the Coulomb logarithm is modified to ${ \kmax \simeq \min\{ \mb\sigma/\hbar , \sigma^2/[G(\mt+\mb)] \} }$.
The similarity with Eq.~\eqref{eq:landau_c} is striking.
The main difference arises in the diffusion term,
where the mass of the background classical particles $m$
is now replaced with ${ [\mb \!+\! (h/\mb)^{3} \Fb (\bvp)] }$.

\subsection{Relaxation of waves by waves}
\label{sec:WaveWithWave}

The Landau equation can also be generalized to describe
the self-consistent evolution of a halo \DF\ composed of ultra-light fuzzy particles.
In this case, Eq.~\eqref{eq:landau_b} becomes
\begin{equation}
  \label{eq:landau_fdm}
  \frac{\p \Fb(\bv)}{\p t} = 2 G^2 \ln\Lambda \frac{\p }{\p v_i} \!\int\!\! \rd \bvp \, u_{ij}(\bv - \bvp) 
  \bigg\{
  \bigg[\mb + \frac{h^3}{\mb^3}\Fb(\bvp)\bigg]\Fb(\bvp) \frac{\p \Fb(\bv)}{\p v_j}
  - \bigg[\mb + \frac{h^3}{\mb^3}\Fb(\bv)\bigg]
\, \frac{\p \Fb (\bvp)}{\p \vp_{j}} \, \Fb(\bv)
  \bigg\}.
\end{equation}
Once again, the similarity with Eq.~\eqref{eq:landau_b} is striking.
The main difference arises in the friction component,
where $\mt$, the mass of the test particle,
is now replaced with ${ [\mb \!+\! (h/\mb)^{3} \Fb (\bv)] }$ (see also~\citealt{Lancaster+2020} for an extensive discussion of dynamical friction in FDM halos).
Deriving Eq.~\eqref{eq:landau_fdm} is one of the main goals of the present paper.
We give a physically motivated derivation of this result relying on the \BNUU\@ equation in Section \ref{sec:DerivationBNUU}, 
and a more rigorous derivation starting from the quasi-linear expansion of the \SP\ system in Section \ref{sec:DerivationKlim}.
Before that, let us first discuss
some of the main properties of Eq.~\eqref{eq:landau_fdm}.

\subsection{Some properties of the Landau equation}
\label{sec:Properties}

First, we note that Eq.~\eqref{eq:landau_fdm}, like  Eqs.~\eqref{eq:landau_c} and~\eqref{eq:landau_b}, is a flux-conservative equation,
that is, it has the form ${\p \Fb / \p t = - \p F_i/\p v_i }$,
with $F_i$ the mass flux in direction $v_{i}$.
Moreover, Eq.~\eqref{eq:landau_fdm} can be written as a \FP\ equation
\begin{equation}
  \label{eq:fp_fdm}
  \frac{\p \Fb(\bv)}{\p t} =
  - \frac{\p }{\p v_i} \left[ D_i(\bv)\Fb(\bv) \right]
  + \frac{1}{2} \frac{\p^{2} }{\p v_{i} \p v_{j}} \left[D_{ij}(\bv)\Fb(\bv) \right] ,
\end{equation}
where the flux is
\begin{equation}
 F_i=-D_i(\bv)\Fb(\bv)+\frac{1}{2}\frac{\p}{\p v_j}\left[D_{ij}(\bv)\Fb(\bv)\right].
\end{equation}
The first- and second-order diffusion coefficients
are given by
\begin{equation}
  \label{eq:D_cb}
  D_{i}(\bv) = \Dc_{i} (\bv) + \Db_{i} (\bv) ; \quad
  D_{ij}(\bv) = \Dc_{ij} (\bv) + \Db_{ij} (\bv).
\end{equation}
Here, ${ \Dc_{i} (\bv) }$ and ${ \Dc_{ij} (\bv) }$ are the classical diffusion coefficients,
while $\Db_{i}$ and $\Db_{ij}$ (``$\mathrm{b}$'' for boson)
capture the contributions associated with wave interference.
The first-order or ``drift" coefficients read 
\begin{align}
\Dc_{i} (\bv) {} & = 4 G^{2} \mb \ln \Lambda \!\! \int \!\! \rd \bvp \, u_{ij} (\bv - \bvp) \, \frac{\p \Fb (\bvp)}{\p v_{j}^{\prime}} ,
\nonumber
\\
\Db_{i} (\bv) {} & = 2 G^{2} \frac{h^{3}}{\mb^{3}} \ln \Lambda \!\! \int \!\! \rd \bvp \, u_{ij} (\bv - \bvp) \, \bigg[ \Fb (\bv) \, \frac{\p \Fb (\bvp)}{\p v_{j}^{\prime}} + \frac{\p \Fb^{2} (\bvp)}{\p v_{j}^{\prime}} \bigg] .
\label{Di_init}
\end{align}
while the second-order or ``diffusion" coefficients are
\begin{align}
\Dc_{ij} (\bv) {} & = 4 G^{2} \mb \ln \Lambda \!\! \int \!\! \rd \bvp \, u_{ij} (\bv - \bvp) \, \Fb (\bvp) ,
\nonumber
\\
\Db_{ij} (\bv) {} & = 4 G^{2} \, \frac{h^{3}}{\mb^{3}} \, \ln \Lambda \!\! \int \!\! \rd \bvp \, u_{ij} (\bv - \bvp) \, \Fb^{2} (\bvp) .
\label{Dij_init}
\end{align}
We can rewrite these coefficients in terms of Rosenbluth potentials~\citep{Rosenbluth1957}. We use the relations
\begin{equation}
u_{ij} (\bv - \bvp) = \pi \frac{\p^{2}}{\p v_{i} \p v_{j}} |\bv - \bvp| ;
\quad
\frac{\p }{\p v_{j}} u_{ij} (\bv - \bvp) = 2 \pi \frac{\p }{\p v_{i}} \frac{1}{|\bv - \bvp|} .
\label{Rosenbluth_relations}
\end{equation}
Equation~\eqref{Di_init} then becomes
\begin{align}
\Dc_{i} (\bv) {} & = 8 \pi G^{2} \mb \ln \Lambda \, \frac{\p }{\p v_{i}} \! \int \!\! \rd \bvp \, \frac{\Fb (\bvp)}{|\bv - \bvp|} ,
\nonumber
\\
\Db_{i} (\bv) {} & = 4 \pi G^{2} \frac{h^{3}}{\mb^{3}} \ln \Lambda \bigg[ \frac{\p }{\p v_{i}} \!\! \int \!\! \rd \bvp \, \frac{\Fb^{2} (\bvp)}{|\bv - \bvp|} + \Fb (\bv) \frac{\p }{\p v_{i}} \!\! \int \!\! \rd \bvp \, \frac{\Fb (\bvp)}{|\bv - \bvp|} \bigg]
\nonumber
\\
{} & = 4 \pi G^{2} \meff \ln \Lambda \bigg[ \frac{\p }{\p v_{i}} \bigg[ \! \int \!\! \rd \bvp \, \frac{\Feff (\bvp)}{|\bv - \bvp|} \bigg] + \frac{\Feff (\bv)}{\Fb (\bv)} \frac{\p }{\p v_{i}} \bigg[ \! \int \!\! \rd \bvp \, \frac{\Fb (\bvp)}{|\bv - \bvp|} \bigg] \bigg] .
\label{Di_Rosenbluth}
\end{align}
while Eq.~\eqref{Dij_init} becomes
\begin{align}
\Dc_{ij} (\bv) {} & = 4 \pi G^{2} \mb \ln \Lambda \, \frac{\p^{2} }{\p v_{i} \p v_{j}} \! \int \!\! \rd \bvp \, |\bv - \bvp| \, \Fb (\bvp) ,
\nonumber
\\
\Db_{ij} (\bv) {} & = 4 \pi G^{2} \frac{h^{3}}{\mb^{3}} \ln \Lambda \, \frac{\p^{2} }{\p v_{i} \p v_{j}} \! \int \!\! \rd \bvp \, |\bv - \bvp| \, \Fb^{2} (\bvp) 
\nonumber
\\
{} & = 4 \pi G^{2} \meff \ln \Lambda \, \frac{\p^{2} }{\p v_{i} \p v_{j}} \! \int \!\! \rd \bvp \, |\bv - \bvp| \, \Feff (\bvp) .
\label{Dij_Rosenbluth}
\end{align}
In these expressions, we introduced in particular an effective \DF\@, ${ \Feff (\bv) }$, and an effective mass, $\meff$, through
\begin{equation}
  \label{def_eff}
  \Feff (\bv) = \frac{\!\int\! \rd \bvp \, \Fb(\bvp)}{\!\int\! \rd \bvp \, \Fb^2(\bvp)}\Fb^2(\bv);
  \quad \meff = \frac{h^3}{\mb^3}\frac{\!\int\! \rd \bv \, \Fb^2(\bv)}{\!\int\! \rd \bv \, \Fb(\bv)} .
\end{equation}

As already discussed in Paper I and recovered in Eq.~\eqref{Dij_Rosenbluth},
the fuzzy diffusion term, ${ \Db_{ij} (\bv) }$, is identical to the classical one, ${ \Dc_{ij} (\bv) }$, except that the mass of the particle and the \DF\ are replaced by
their effective counterparts $\meff$ and $\Feff$.
In particular, when the underlying \DF\ is Maxwellian, these diffusion coefficients are the same as those in a halo of classical particles with mass
${ \meff =  \rho_0 {(\lambda_\sigma/\sqrt{4\pi})}^3 }$ and effective velocity
dispersion ${ \sigmaeff = \sigma/\sqrt{2} }$.
As a result, in the fuzzy case,
the relaxation time from Eq.~\eqref{eq:intro_I} becomes
\begin{equation}
  \label{eq:trlx}
  \trelax \simeq 0.34 \frac{\sigmaeff^3}{G^2\meff\rho_0\ln\Lambda} =
  0.34\frac{\mb^3\sigma^6}{G^2 {(2\pi)}^{3/2} \hbar^3\rho_0^2\ln\Lambda}.
\end{equation}

When the \DF\ is isotropic, i.e., when ${ \Fb(\bv) }$ depends only on ${ v = |\bv| }$, Eq.~\eqref{eq:landau_fdm} can be rewritten in an even simpler form.
To do so, we rely once again on the properties of the Rosenbluth potentials,
see Eqs.~{(L22)} and~{(L23)} of~\cite{Binney+2008}.
In particular, we can write
\begin{equation}
\!\! \int \!\! \rd \hbvp \, u_{ij} (\bv - \bvp) = 
\begin{cases}
\displaystyle 4 \pi^{2} \bigg( \frac{\delta_{ij}}{v} - \frac{v_{i}v_{j}}{v^{3}} - \frac{v^{\prime 2} \delta_{ij}}{3 v^{3}} + \frac{v^{\prime 2} v_{i} v_{j}}{v^{5}} \bigg) 
& \text{ for } \quad \vp < v ,
\\[2.0ex]
\displaystyle \frac{8 \pi^{2}}{3} \frac{\delta_{ij}}{\vp} 
& \text{ for } \quad v < \vp ,
\end{cases}
\label{int_iso_1}
\end{equation}
as well as
\begin{equation}
\!\! \int \!\! \rd \hbvp \, u_{ij} (\bv - \bvp) \, \vp_{j} = 
\begin{cases}
\displaystyle \frac{8 \pi^{2}}{3} \frac{v^{\prime 2} v_{i}}{v^{3}} 
& \text{ for } \quad \vp < v ,
\\[2.0ex]
\displaystyle \frac{8 \pi^{2}}{3} \frac{v_{i}}{\vp}
& \text{ for } \quad v < \vp ,
\end{cases}
\label{int_iso_2}
\end{equation}
where we recall that the sum over $j$ is implied
in the second equation.
In order to get that last relation, we used the fact that $u_{ij} (\bv) v_{j}=0$ in conjunction with Eq.~\eqref{int_iso_1}.

Then, after some lengthy manipulations, Eq.~\eqref{eq:landau_fdm} becomes
\begin{align}
  \label{eq:landau_fdm_iso}
  \frac{\p \Fb(v)}{\p t}
  = {} &
    \frac{16\pi^2 G^2\ln\Lambda}{3} \frac{1}{v^{2}}
    \frac{\rd}{\rd v} \bigg\{ \frac{1}{v}\frac{\rd \Fb(v)}{\rd v} \!\!\int_{0}^{v}\!\! \rd \vp \,  
    v^{\prime 4} \Fb(\vp)\bigg[ \mb +
    \frac{h^3}{\mb^3}\Fb(\vp) \bigg]
    \nonumber
    \\
    {} & \quad +v^2\frac{\rd \Fb(v)}{\rd v} \!\!\int_{v}^{\infty}\!\! \rd \vp \, 
    \, \vp \Fb(\vp) \bigg[ \mb +
    \frac{h^3}{\mb^3}\Fb(\vp) \bigg]
    + 3\Fb(v)\bigg[ \mb +\frac{h^3}{\mb^3}\Fb(v) \bigg]
    \!\!\int_{0}^{v}\!\! \rd \vp \, v^{\prime 2} \Fb(\vp)
    \bigg\} .
\end{align}

In the limit of high phase-space density, i.e., many particles per unit phase-space cell of volume $h^3$, or
\begin{equation}
\label{eq:high}
\frac{h^{3}}{\mb^{4}} \Fb \gg 1,\quad\mbox{or}\quad \meff \simeq \rho_{0} \bigg( \frac{h}{\mb \sigma} \bigg)^{3} \gg \mb ,
\end{equation}
we can reduce Eqs.~\eqref{eq:landau_fdm} and \eqref{eq:landau_fdm_iso} to
\begin{align}
  \frac{\p \Fb(\bv)}{\p t} {} & = 2 G^{2} \ln \Lambda \frac{h^{3}}{\mb^{3}} \frac{\p}{\p v_i}
  \!\!\int\!\! \rd \bvp \, u_{ij} (\bv - \bvp) 
  \bigg[\Fb^2(\bvp) \frac{\p \Fb(\bv)}{\p v_j}
  - \Fb^2(\bv) \frac{\p \Fb(\bvp)}{\p \vp_{j}} \bigg] 
  \\
  {} & = \frac{16\pi^2 G^2\ln\Lambda}{3}
    \frac{h^3}{\mb^3} \frac{1}{v^{2}} \frac{\rd}{\rd v}
    \bigg[\frac{1}{v}\frac{\rd\Fb(v)}{\rd v}
    \!\!\int_{0}^{v}\!\! \rd \vp \, v^{\prime 4} \Fb^2(\vp) +v^2
    \frac{\rd \Fb(v)}{\rd v}
    \!\!\int_{v}^{+\infty}\!\! \rd \vp \, 
    \vp \Fb^2(\vp)+3\Fb^2(v)
    \!\!\int_{0}^{v}\!\! \rd \vp \, v^{\prime 2} \Fb(\vp) \bigg].
    \nonumber 
\end{align}
The second of these is equivalent to Eq.~{(S23)} in~\cite{Levkov+2018}.

The generic steady state of Eq.~\eqref{eq:landau_fdm_iso}
is the Bose-Einstein \DF\@,
\begin{equation}
  \label{eq:bes}
  \Fb(v) = \frac{\mb^4}{h^3}\frac{1}{\re^{\frac{1}{2}\beta v^2}/z-1}.
\end{equation}
In that expression, the inverse temperature $\beta$
and the fugacity $z$ are determined from the conservation of mass
and kinetic energy per unit volume,
following Eq.~\eqref{def_rho}.
In practice, one has
\begin{equation}
  \label{eq:bes_rho0}
  \rho_0 = \frac{\mb^4{(2\pi)}^{3/2}}{h^3\beta^{3/2}}\Li_{3/2}(z)
  ; \quad
  \rho_0\sigma^2 = \frac{\mb^4{(2\pi)}^{3/2}}{h^3\beta^{5/2}}\Li_{5/2}(z),
\end{equation}
with ${ \Li_n(z) = \sum_{k=1}^{+\infty} z^{k}/k^{n} }$ the polylogarithm.
Combining these equations to eliminate $\beta$, we can write
\begin{equation}
  \label{eq:rho-sigma}
  \sigma^3 = \frac{h^3\rho_0}{\mb^4{(2\pi)}^{3/2}}\frac{\Li_{5/2}^{3/2}(z)}{\Li_{3/2}^{5/2}(z)},
\end{equation}
which has a minimum value, $\sigmac$, reached at $z=1$, 
\begin{align}
  \label{eq:rho-sigma-minimum}
  \sigmac^3 {} & =
  \frac{\zeta^{3/2}(5/2)}{\zeta^{5/2}(3/2)} \frac{h^3\rho_0}{\mb^4{(2\pi)}^{3/2}}
  \nonumber
  \\
  {} & \simeq 1.09 \, \big( 100 \kms \big)^{3} \frac{\rho_0}{0.01 \Msun
    \pc^{-3}} \bigg( \frac{\mb}{20 \eV} \bigg)^{-4} ,
\end{align}
with ${ \zeta(x)=\Li_{x} (1) }$ the Riemann zeta function.
The corresponding value of $\beta$ is then given by
\begin{equation}
\betac = \frac{\zeta(5/2)}{\zeta(3/2)} \frac{1}{\sigmac^{2}} = \bigg[\frac{\mb \zeta(3/2)}{\rho_0} \bigg]^{2/3} \, \frac{\mb^{2}}{2 \pi \hbar^{2}} .
\label{def_betac}
\end{equation}
 
Systems with an initial velocity dispersion below $\sigmac$ cannot occupy the steady state given by Eq.~\eqref{eq:bes}. In this case the steady state is given by
\begin{equation}
  \label{eq:bes_steady}
  \Fb(v) = \frac{\mb^4}{h^3}\frac{1}{\re^{\frac{1}{2}\beta v^2}-1} + \rho_\rs \, \deltaD (\bv),
\end{equation}
where the last term is the Bose--Einstein condensate or soliton\footnote{For the typical boson masses $\mb=10^{-21} \text{--} 10^{-22}\eV$ suggested for FDM, $\sigmac$ is much larger than galaxy velocity dispersions and the thermal equilibrium state is always a condensate.}.
In that expression, the inverse temperature follows from the second of Eqs.~\eqref{eq:bes_rho0} and reads
\begin{equation}
\beta^{5/2} = \frac{\mb^{4} (2 \pi)^{3/2}}{h^{3} \rho_{0} \sigma^{2}} \, \zeta(5/2) .
\label{eq:beta_bes}
\end{equation}
Finally, the mass density in the soliton is determined from the conservation of mass and energy.
As such, it is given by
\begin{equation}
  \label{eq:rho_s}
\rho_\rs = \bigg(1-\frac{\betac^{3/2}}{\beta^{3/2}}\bigg)\rho_0=\bigg(1 - \frac{\sigma^{6/5}}{\sigmac^{6/5}}\bigg)\rho_0 .
\end{equation}
As stated in Section \ref{sec:intro}, we do not consider the formation of the condensate or soliton in this paper.

\bigskip

\section{Deriving the kinetic equation from the Boltzmann--Nordheim--Uehling--Uhlenbeck (BNUU) equation}
\label{sec:DerivationBNUU}

In this section we give a (relatively) simple derivation
of Eq.~\eqref{eq:landau_fdm}.
The \BNUU\ equation~\citep{nord28,uu33,Erdos2004} is a heuristic generalization of the Boltzmann equation to quantum systems, see also. For a homogeneous system the BNUU equation reads
\begin{align}
\frac{\p f(\bp_1)}{\p t} {} & = \!\!\int\!\! \rd \bp_2 \rd \bp_3 \rd \bp_4 \, 
S^{(4)}(\bp_1,\bp_2,\bp_3,\bp_4)\big\{f(\bp_3)f(\bp_4)[1+\eps h^3
f(\bp_1)][1+\eps h^3 f(\bp_2)]
\nonumber
\\
{} &
\qquad  
- f(\bp_1)f(\bp_2)[1+\eps h^3 f(\bp_3)][1+\eps h^3 f(\bp_4)]
\big\},
\label{eq:bnuu}
\end{align}
with ${ \eps = 0 }$ for a classical system, ${ +1 }$ for bosons, and ${ -1 }$ for fermions. 
Here $\bp$ is the momentum and $f(\bp)\rd\br\rd\bp$ is the number of particles in a phase-space volume element $\rd\br\rd\bp$ (the dependence of the \DF\ ${ f(\bp) }$ on time is not shown explicitly). The function $S^{(4)}$ describes the rate at which particles with momenta $\bp_1$ and $\bp_2$ are scattered to momenta $\bp_3$ and $\bp_4$. More precisely,  $S^{(4)}(\bp_1 , \bp_2 , \bp_3 , \bp_4) f(\bp_1)f(\bp_2) \rd\bp_1\rd\bp_2 \rd\bp_3 \rd\bp_4$ is the rate per unit volume at which particles are scattered from the momentum-space volumes $\rd\bp_1$ and $\rd\bp_2$ into the volumes $\rd\bp_3$ and $\rd\bp_4$.
For ${ \eps = 0 }$, Eq.~\eqref{eq:bnuu} reduces to the classical Boltzmann equation, and for ${ \eps = -1 }$ factors such as ${ 1-h^3f(\bp_3) }$ ensure that no particles are scattered into states that are fully occupied according to the Pauli principle.

Since momentum is conserved in collisions, the function $S^{(4)}$ must contain a factor $\deltaD (\bp_3+\bp_4-\bp_1-\bp_2)$ and we use this to carry out the integral over $\bp_4$. Then we replace $\bp_3$ by the momentum transfer ${ \bq = \bp_3 - \bp_1 }$ to obtain
\begin{align}
\frac{\p f(\bp_1)}{\p t} {} & = \!\!\int\!\! \rd \bp_2 \rd \bq \, S(\bp_1+\half \bq , \bp_2 - \half \bq ;\bq)\big\{f(\bp_1+\bq)f(\bp_2-\bq)[1+\eps h^3 f(\bp_1)][1+\eps h^3 f(\bp_2)] \nonumber
\\
{} & \qquad  -f(\bp_1)f(\bp_2)[1+\eps h^3 f(\bp_1+\bq)][1+\eps h^3 f(\bp_2-\bq)]\big\}.
\label{eq:uu3}
\end{align}
Here we have rewritten $S^{(4)}$ as a function of three variables, $S(\bp_1+\half\bq,\bp_2-\half\bq;\bq) \equiv S^{(4)}(\bp_1,\bp_2,\bp_1+\bq,\bp_4) \, \deltaD(\bp_4+\bq-\bp_2)$. For gravitational scattering between particles of mass $\mb$, $S$ is given by \citep[e.g.,][]{goodman83}
\begin{equation}
\label{eq:xsect}
  S(\ba , \bb ; \bq) = \frac{4G^2\mb^5}{q^4}\,\deltaD [(\bb-\ba) \!\cdot\! \bq] .
\end{equation}
In that expression, the delta function ensures that the relative momentum is conserved in the collision, ${ |\bp_4 \!-\! \bp_3| = |\bp_2 \!-\! \bp_1| }$, and the factor $q^{-4}$ reflects the angular dependence of the Coulomb differential scattering cross-section, ${ |\sin^2(\theta/2)|^{-4} }$,
where $\theta$ is the scattering angle.

We now assume that $\bq$ is small
(equivalent to the \FP\ approximation of weak deflections)
and that $S$ varies slowly with $\bq$ in its first two arguments.
This allows us then to expand Eq.~\eqref{eq:uu3} to second order in $\bq$.
 We abbreviate the notation by writing ${ f_1=f(\bp_1) }$, ${ \p_t f=\p f/\p t }$, ${ \p_{1,i} f = \p f/\p p_{1i} }$, ${ \p_{1,ij}f=\p^2 f/\p p_{1i}\p p_{1j} }$, etc.
 We also assume summation over repeated indices. Then
\begin{align}
\label{eq:calc_dfdt}
\p_t f (\bp_1) {} & = \!\!\int\!\! \rd \bp_2 \rd \bq \, 
\big( S +\half q_i\p_{1,i} S - \half q_i\p_{2,i}S\big)
\nonumber
\\
{} & \quad \times \bigg[(f_1+q_i\p_{1,i}f_{1}+\half q_iq_j\p_{1,ij}f_{2})(f_2- q_k\p_{2,k}f_{2}+\half q_kq_m \p_{2,km}f_{2})
(1+\eps h^3 f_1)(1+\eps h^3 f_2)
\nonumber
\\
{} & \quad - f_1f_2(1+\eps h^3 f_1 +\eps h^3 q_i\p_{1,i}f_{1} + \half\eps h^3 q_iq_j \p_{1,ij}f_{1})(1+\eps h^3 f_2 -\eps h^3 q_k\p_{2,k} f_{2}+\half\eps h^3 q_kq_m \p_{2,km}f_{2}) \bigg] ;
\end{align}
here the function ${ S (\bp_{1} + \half \bq, \bp_{2} - \half \bq;\bq) }$
was only expanded to first order in $\bq$,
as the terms within brackets vanish at zero order in $\bq$.
In that expression,
$S$ stands for $S(\bp_1,\bp_2;\bq)$ which is even in $\bq$ by Eq.~\eqref{eq:xsect} (or more generally by detailed balance).
As a result, the terms that are first order in $\bq$ vanish when integrated.
Thus, Eq.~\eqref{eq:calc_dfdt} simplifies to
\begin{align}
\p_t f(\bp_1) {} & = \!\!\int\!\! \rd \bp_2 \rd \bq \, q_i q_j \big\{ \half S[\p_{1,ij}f_{1} f_2(1+\eps h^3 f_2) + \p_{2,ij}f_{2} f_1(1+\eps h^3 f_1) -2\p_{1,i}f_{1}\,\p_{2,j}f_{2}(1+\eps h^3 f_1+\eps h^3 f_2)]
\nonumber
\\
{} & \quad +\half (\p_{1,i}S-\p_{2,i} S)[\p_{1,j}f_{1} f_2(1+\eps h^3 f_2) - \p_{2,j}f_{2} f_{1} (1+\eps h^3 f_1)]\big\}.
\end{align}
Integrating the term involving $\p_{2,i}S$ by parts gives
\begin{align}
\p_t f(\bp_1) {} & = \!\!\int\!\! \rd \bp_2 \rd \bq \, q_i q_j \big \{ \half S[\p_{1,ij}f_{1} f_2(1+\eps h^3 f_2)  -\p_{1,i}f_{1}\,\p_{2,j}f_{2} (1+2\eps h^3 f_1)]
\nonumber
\\
{} & \quad +\half \p_{1,i}S[\p_{1,j}f_{1} f_2(1+\eps h^3 f_2) - \p_{2,j}f_{2} f_1(1+\eps h^3 f_1)]\big\}
\nonumber
\\
{} & = \half \p_{1,i} \!\!\int\!\! \rd \bp_2 \rd \bq \, q_i q_j S[\p_{1,j}f_{1} f_2(1+\eps h^3 f_2) - \p_{2,j}f_{2} f_1(1+\eps h^3 f_1)]
\end{align}
Following Eq.~\eqref{eq:xsect}, the integral over $\bq$ yields
\begin{equation}
  \!\!\int\!\! \rd \bq \, q_i q_j S(\bp_1,\bp_2;\bq)=4G^2\mb^5\ln\Lambda\, u_{ij}(\bp_1-\bp_2) ,
\end{equation}
where $u_{ij}$ is the collision kernel from Eq.~\eqref{eq:uij_I},
and ${ \Lambda = \qmax/\qmin }$ with $\qmax$ and $\qmin$
the maximum and minimum impact parameters included in the integral.

Now change variables from momentum $\bp$ to velocity ${ \bv = \bp/\mb }$, from $\bv_1$ and $\bv_2$ to $\bv$ and $\bvp$, and from the number density in position-momentum phase space ${ f(\bp) }$ to the mass density in position-velocity space ${ \Fb(\bv)=\mb^{4} f(\mb\bv) }$. We find
\begin{align}
\label{eq:Landau_BNUU}
\frac{\p\Fb(\bv)}{\p t} {} & = 2G^2\ln\Lambda \frac{\p}{\p v_i} \!\!\int\!\! \rd \bvp \, u_{ij}(\bv-\bvp) \bigg\{ \bigg[\mb+\eps \frac{h^3}{\mb^3}\Fb(\bvp)\bigg]\Fb(\bvp) \frac{\p\Fb(\bv)}{\p v_{j}}
-\bigg[1+\eps\frac{h^3}{\mb^3}\Fb(\bv)\bigg]\frac{\p \Fb(\bvp)}{\p \vp_{j}} \Fb(\bv)\bigg\} .
\end{align}
which proves that Eq.~\eqref{eq:landau_fdm} follows directly from the BNUU equation.
Of course, the present derivation remains heuristic
because it stems from the heuristic BNUU equation (\ref{eq:bnuu}).
In the following section, we will present a more careful  derivation of Eq.~\eqref{eq:landau_fdm}, through
a detailed study of the self-consistent dynamics
of the wavefunction of an FDM halo.

\bigskip

\section{Deriving the kinetic equation from the Schr\"{o}dinger--Poisson equation}
\label{sec:DerivationKlim}

We now revisit the derivation of the kinetic equation~\eqref{eq:landau_fdm}
by studying the gravitational interaction of waves ${ \psi (\br , t) }$
that evolve according to the \SP\ equations~\citep{Ruffini+1969}
\begin{align}
  \label{eq:schr-poisson}
  \ri \hbar \, \frac{\p \psi(\br, t)}{\p t} 
  {} & 
    = -\frac{\hbar^2}{2\mb} \nabla^{2} \psi(\br, t) + \mb \Phi(\br, t) \psi(\br,t),
    \\
  \label{eq:pe}
  \nabla^{2} \Phi(\br, t) 
  {} &
    = 4\pi G ({|\psi(\br, t)|}^2 - \rho_0).
\end{align}
In the Poisson equation, the term $-\rho_0$ appears because of the Jeans swindle,
already described in Section~\ref{sec:KineticEquations}.
We also note that the wavefunction ${ \psi(\br,t) }$ is normalized so that ${ |\psi(\br,t)|^2 }$ is a mass density,
i.e.,\ one has ${ \langle | \psi (\br , t) |^{2} \rangle = \rho_{0} }$.

The Wigner function~\citep{Wigner1932} 
\begin{align}
  \label{eq:wigner_I}
  W(\br,\bv, t) 
  {} & 
    = \!\!\int\!\! \frac{\rd \bs}{(2 \pi)^{3}} \, \re^{-\ri \bv \cdot \bs} \, \psi\big(\br +\half\hbar\bs/ \mb, t \big) \, \psi^*\big(\br - \half\hbar\bs/ \mb, t
    \big)
\end{align}
is the quantum analog to the discrete classical phase-space \DF\@, ${ \Fd (\br,\bv,t) =
\mb\sum_{i=1}^N \deltaD [\br - \br_{i} (t)] \, \deltaD [\bv - \bv_{i} (t)] }$.
As such, ${ W (\br , \bv , t) }$ can be interpreted classically as the density of bosons
in the infinitesimal volume ${ \rd \br \rd \bv }$.
When averaged over realisations,
${ \oW = \langle W \rangle }$ is the quantum analog
to the mean phase-space
\DF\@, ${ \Fb = \langle \Fd \rangle }$,
i.e., the \DF\@ whose kinetic equation we want to derive.
With our conventions, we also note that $W$ and $\oW$ satisfy the normalisations
\begin{equation}
    \!\!\int\!\! \rd\bv \, W(\br,\bv,t) ={|\psi(\br,t)|}^2; \quad \!\!\int\!\! \rd \bv \, \oW (\bv) = \rho_0 ,
        \label{eq:wigint}
\end{equation}
where we used the relation ${ \!\int\! \rd \bv/(2 \pi)^{3} \re^{- \ri \bv \cdot \bs} \, \!=\! \deltaD (\bs) }$.

The time evolution of ${ W (\br , \bv , t) }$
follows from the Schr\"{o}dinger equation~\eqref{eq:schr-poisson}.
More precisely, using the relation
\begin{equation}
\nabla^{2} \psi \big( \br \pm \half \hbar \bs / \mb , t \big) = \pm \frac{2 \mb}{\hbar} \frac{\p }{\p \bs} \!\cdot\! \bigg[ \bnabla \psi \big( \br \pm \half \hbar \bs / \mb , t \big) \bigg] 
\label{transformation_coordinates}
\end{equation}
to perform an integration by parts with respect to ${ \rd \bs }$
in Eq.~\eqref{eq:wigner_I},
we obtain
\begin{align}
  \frac{\p W(\br,\bv, t)}{\p t} + \bv \!\cdot\! \bnabla W(\br,\bv, t) =
    \ri \frac{\mb}{\hbar} \!\!\int\!\! \frac{\rd \bs}{(2 \pi)^{3}} \, \re^{-\ri \bv \cdot \bs} \,
    \bigg[
    {} &
    \Phi\big(\br - \half \hbar\bs/\mb, t \big)
    - \Phi\big(\br + \half\hbar\bs/\mb, t \big)
    \bigg]
    \nonumber
    \\
    \times
    {} &
    \psi\big(\br +
    \half\hbar\bs/\mb, t \big) \, \psi^*\big(\br - \half\hbar\bs/\mb, t
    \big) .
    \label{eq:wigner_eom_long}
\end{align}

To better emphasize the analogy with the classical case,
we can rewrite Eq.~\eqref{eq:wigner_eom_long} in the shorter form
\begin{equation}
\frac{\p W(\br,\bv,t)}{\p t} + \bv \!\cdot\! \bnabla W (\br , \bv , t) = \oD \big[ \Phi(\br,t) , W(\br,\bv,t) \big] ,
\label{eq:wigner_eom}
\end{equation}
where the non-linear term from the r.h.s.\ reads
\begin{align}
  \label{eq:D_op}
  \oD \big[ \Phi(\br,t) , W(\br,\bv,t) \big] \equiv
  \ri \!\!\int\! \rd \bk \, \!\!\int_{\mB} \frac{\rd \omega}{2 \pi} \, \re^{\ri (\bk \cdot \br - \omega t)} \,
  \hPhi (\bk,\omega) \, 
  \hoD_\bk \big[ W(\br,\bv,t) \big] .
\end{align}
In this expression, the Laplace-Fourier transformed potential, ${ \hPhi (\bk , \omega) }$, is introduced with the convention 
\begin{equation}
\hPhi (\bk , \omega) = \!\! \int \!\! \frac{\rd \br}{(2 \pi)^{3}} \!\! \int_{0}^{\infty} \!\!\!\! \rd t \, \re^{- \ri (\bk \cdot \br - \omega t)} \, \Phi (\br , t);
\quad
\Phi (\br , t) = \!\! \int \!\! \rd \bk \!\! \int_{\mB}  \frac{\rd \omega}{2 \pi} \, \re^{\ri (\bk \cdot \br - \omega t)} \, \hPhi (\bk , \omega) ,
\label{def_hPhi}
\end{equation}
where the Bromwich contour, $\mB$, has to pass above all the poles of the integrand,
i.e.,\ ${ \ImPart (\omega) }$ has to be large enough.
In Eq.~\eqref{eq:D_op}, we also defined the finite-difference operator 
\begin{equation}
  \label{eq:Dhat}
  \hoD_\bk \big[ W(\bv) \big]  = 
  \frac{\mb}{\hbar} \bigg[ W \bigg(\! \bv+\frac{\hbar}{2\mb}\bk \!\bigg) -
  W\bigg(\! \bv-\frac{\hbar}{2\mb}\bk \!\bigg)\bigg].
  \end{equation}
We already note that in the classical limit,
${ \hbar \to 0 }$, this finite-difference operator satisfies
\begin{equation}
\label{classical_derivatives}
\hoD_\bk \big[ W(\br,\bv,t) \big] \to \bk \!\cdot\! \frac{\p W(\br,\bv,t)}{\p \bv};
\quad \oD \big[ \Phi(\br,t),W(\br,\bv,t) \big] \to \bnabla \Phi(\br,t) \!\cdot\! \frac{\p W(\br,\bv,t)}{\p \bv} ;
\end{equation}
so that Eq.~\eqref{eq:wigner_eom} reduces
to the classical collisionless Boltzmann equation~\citep{Binney+2008}
\begin{equation}
\frac{\p W}{\p t} + \bv \!\cdot\! \bnabla W - \bnabla \Phi \!\cdot\! \frac{\p W}{\p \bv} = 0 .
\label{classical_CBE}
\end{equation}

Our goal now is to describe how the unavoidable quantum fluctuations in the system lead to the relaxation of the halo's underlying mean \DF\@.
To pursue this, we perform a quasi-linear expansion of Eq.~\eqref{eq:wigner_eom}. This is a standard procedure in kinetic theory
and we generally follow the method presented in~\cite{Chavanis2012Historic}
in the context of the relaxation of classical discrete self-gravitating systems.

We write the Wigner distribution as a perturbation around its ensemble
average,
\begin{equation}
W = \oW + f .
\label{expansion_W}
\end{equation}
Since we used the Jeans swindle in Eq.~\eqref{eq:pe},
we note that ${ \Phi(\br,t) }$ is already the fluctuating potential.
Separating the average and fluctuating components of Eq.~\eqref{eq:wigner_eom},
we obtain two evolution equations for the instantaneous fluctuations, ${ \p f / \p t }$,
and the system's mean \DF\@, ${ \p \oW / \p t }$.
They read
\begin{align}
  \label{eq:wigner_eom_f}
  \frac{\p f(\br,\bv, t)}{\p t} + \bv \!\cdot\! \bnabla f(\br,\bv, t) 
  {} &
    =
    \oD\big[ \Phi(\br,t) , \oW (\br,\bv,t) \big] ,
    \\
  \label{eq:Wbar_eom}
  \frac{\p \oW (\br,\bv, t)}{\p t} + \bv \!\cdot\! \bnabla \oW (\br,\bv, t) 
  {} &
    =
    \big\langle \oD \big[ \Phi(\br,t) , f(\br,\bv,t) \big] \big\rangle ,
\end{align}
where Eq.~\eqref{eq:wigner_eom_f} has been linearized in the perturbations by neglecting the quadratic term
${ \oD[\Phi(\br,t),f(\br,\bv,t)] }$ therein.
These equations are valid in the weak-coupling limit where ${ \td \ll \trelax }$. In the present case, the dynamical time $\td$ and the relaxation time $\trelax$ are approximately
\begin{align}
\td \simeq \max \bigg[ \frac{\hbar}{\mb \sigma^{2}} , (G\rho_{0})^{-1/2} \bigg] ;
\quad
\trelax \simeq \min \bigg[ \frac{\mb^{3} \sigma^{6}}{G^{2} \rho_{0}^{2} \hbar^{3}} , \frac{\sigma^{3}}{G^{2} \mb \rho_{0}} \bigg] .
\label{def_td_trelax}
\end{align}
Here the classical dynamical time is ${ (G\rho_0)^{-1/2} }$ and the classical relaxation time is given by Eq.~\eqref{eq:intro_I}; if the wavelike nature of the particles is important the dynamical time is ${ \sim \lbarsig/\sigma }$, the time taken for a particle traveling at the typical speed to cross a de Broglie wavelength, and the relaxation time is given by Eq.~\eqref{eq:intro_II}.

In the weak-coupling limit, we can assume that ${ \oW }$ is a (quasi)-stationary solution
of the l.h.s.\ of Eq.~\eqref{eq:Wbar_eom},
i.e.,\ we can assume that ${ \oW = \oW (\bv , t) }$
is homogeneous and therefore only slowly changes in time.
The l.h.s.\ of Eq.~\eqref{eq:Wbar_eom} is then the ensemble-averaged
collisionless Boltzmann equation
that describes the free streaming of particles
in the absence of any potential perturbations,
while its r.h.s.\ is the collision term.
In particular, we note that since both the potential ${ \Phi (\br , t) }$
and the perturbed DF ${ f (\br , \bv , t) }$
are quadratic in the wavefunction,
the collision term from Eq.~\eqref{eq:Wbar_eom}
involves a product of four factors ${ \psi (\br , t) }$
evaluated at different locations.

Following these assumptions, we can take the Laplace-Fourier transform
of Eq.~\eqref{eq:wigner_eom_f}, which gives
\begin{equation}
  \label{eq:wigner_eom_lf}
\hf (\bk,\bv, \omega) = - \frac{\hoD_{\bk} \big[ \oW (\bv) \big]}{\omega - \bk \!\cdot\! \bv} \, \hPhi (\bk , \omega) +\ri \frac{\hf_{0} (\bk , \bv)}{\omega - \bk \!\cdot\! \bv} ,
\end{equation}
where the Fourier transforms with respect to space and the Laplace transform  with respect to time
are defined with the conventions of Eq.~\eqref{def_hPhi}.
We also assumed that ${ \oW (\bv) }$ can be taken as constant
on the timescales over which the fluctuations evolve.
In Eq.~\eqref{eq:wigner_eom_lf}, we also introduced
\begin{equation}
\hf_{0} (\bk ,\bv) = \!\! \int \!\! \frac{\rd \br}{(2 \pi)^{3}} \, \re^{- \ri \bk \cdot \br} \, f_{0} (\br , \bv) ,
\label{def_init_Fourier}
\end{equation}
as the Fourier transform of the fluctuations of the DF 
at the initial time, ${ f_{0} (\br , \bv) \!=\! f (\br , \bv , t \!=\! 0) }$.

Using the normalisations from Eq.~\eqref{eq:wigint},
we can rewrite the Poisson equation~\eqref{eq:pe}
in the simple form
\begin{equation}
\nabla^{2} \Phi (\br , t) = 4 \pi G \!\! \int \!\! \rd \bv \, f (\br , \bv , t) ,
\label{rewrite_Poisson}
\end{equation}
which becomes in Fourier-Laplace space
\begin{equation}
  \label{eq:Phi_hat_two}
  \hPhi (\bk,\omega) =   - \frac{4\pi G}{k^2} \!\!\int\!\! \rd \bv \, \hf (\bk, \bv, \omega).
\end{equation}

We now have at our disposal Eqs.~\eqref{eq:wigner_eom_lf} and~\eqref{eq:Phi_hat_two}, which jointly couple the \DF\ and potential fluctuations,
${ \hf (\bk , \bv , \omega) }$ and ${ \hPhi (\bk , \bv , \omega) }$.
Solving these self-consistently amounts then to accounting for collective effects,
i.e.,\ accounting for the ability of the system to amplify its own self-generated perturbations. To make progress, the traditional solution is to act on both sides of Eq.~\eqref{eq:wigner_eom_lf}
with the same operator as in the r.h.s.\ of Eq.~\eqref{eq:Phi_hat_two}.
One immediately obtains
\begin{equation}
\hPhi (\bk , \omega) = \frac{1}{\eps (\bk , \omega)} \, \frac{4 \pi G}{\ri k^{2}} \!\! \int \!\! \rd \bv \, \frac{\hf_{0} (\bk , \bv)}{\omega - \bk \!\cdot\! \bv} ,
\label{eq:Phi_hat0}
\end{equation}
where the dielectric function is
\begin{equation}
\eps(\bk, \omega) = 1
  - \frac{4\pi G}{k^2} \!\!\int\!\! \rd \bv \,
  \frac{ \hoD_\bk \big[ \oW (\bv) \big]}{\omega - \bk \!\cdot\! \bv} .
\label{eq:dielectric}
\end{equation}
As in Eq.~\eqref{classical_derivatives},
one can straightforwardly obtain the classical dielectric function
through the substitution
${ \hoD_\bk [\oW] \to \bk \!\cdot\! \p \oW / \p \bv }$.
As usual, the dielectric function can be rewritten using Landau's prescription.
Following our convention from Eq.~\eqref{def_hPhi},
this amounts to making the replacement ${ \omegaR \to \omegaR \!+\! \ri 0^{+} }$ and using the Plemelj formula
\begin{equation}
\frac{1}{\omegaR \!+\! \ri 0^{+}} = \mP \bigg( \frac{1}{\omegaR} \bigg) - \ri \pi \deltaD (\omegaR) ,
\label{Plemelj_formula}
\end{equation}
for ${ \omegaR \in \mathbb{R} }$ and with ${ \mP }$ being Cauchy's principal value.
We will further discuss the properties of that dielectric function in Section~\ref{sec:Stability},
when investigating the linear stability of the present system.

We can now turn to Eq.~\eqref{eq:Wbar_eom}
in order to relate the long-term evolution of the mean \DF\@,
${ \p \oW (\bv , t) / \p t }$, to the correlations of the initial \DF\ fluctuations,
${ \hf_{0} (\bk , \bv) }$.
Starting from Eq.~\eqref{eq:Wbar_eom},
we can use the definition from Eq.~\eqref{eq:D_op}
to write
\begin{align}
\frac{\p \oW (\bv , t)}{\p t} = \ri \!\! \int \!\! \rd \bk \rd \bkp \, \re^{\ri(\bk+\bkp)\cdot\br}\,\hoD_{\bkp} \bigg[ \!\! \int \!\! \frac{\rd \omega}{2 \pi} \frac{\rd \omegap}{2 \pi} \, \re^{- \ri (\omega + \omegap) t} \, \big\langle \hf (\bk , \bv , \omega) \, \hPhi (\bkp , \omegap) \big\rangle \bigg] ,
\label{eq:D_f0}
\end{align}
In principle the r.h.s.\ depends on position $\br$ but we shall argue below that this dependence vanishes. Since Eq.~\eqref{eq:wigner_eom_lf} has two terms, we get two contributions,
\begin{align}
\frac{\p \oW (\bv , t)}{\p t} = F_{1} (\bv) + F_{2} (\bv) ,
\label{rewrite_average}
\end{align}
where we introduced
\begin{align}
F_{1} (\bv) {} & = - \!\! \int \!\! \rd \bk \rd \bkp \, \re^{\ri(\bk+\bkp)\cdot\br}\,\hoD_{\bkp} \bigg[ \!\! \int \!\! \frac{\rd \omega}{2 \pi} \frac{\rd \omegap}{2 \pi} \, \frac{\re^{- \ri (\omega + \omegap) t}}{\omega - \bk \!\cdot\! \bv} \, \big\langle \hf_{0} (\bk , \bv) \, \hPhi (\bkp , \omegap) \big\rangle \bigg]
\nonumber
\\
F_{2} (\bv) {} & = - \ri \, \!\! \int \!\! \rd \bk \rd \bkp \, \re^{\ri(\bk+\bkp)\cdot\br}\,\hoD_{\bkp} \bigg[ \hoD_{\bk} \big[ \oW (\bv) \big] \!\! \int \!\! \frac{\rd \omega}{2 \pi} \frac{\rd \omegap}{2 \pi} \, \frac{\re^{- \ri (\omega + \omegap)t}}{\omega - \bk \!\cdot\! \bv} \, \big\langle \hPhi (\bk , \omega) \, \hPhi (\bkp , \omegap) \big\rangle \bigg] ,
\label{def_F1_F2}
\end{align}
which, as we will show, respectively capture the contributions from the drift and diffusion components.
To proceed further, we rewrite these two expressions using Eq.~\eqref{eq:Phi_hat0}.
After some manipulations, these two components become
\begin{align}
F_{1} (\bv) {} & =  \ri \!\! \int \!\! \rd \bk \rd \bkp \, \re^{\ri(\bk+\bkp)\cdot\br}\,\hoD_{\bkp} \bigg[ \frac{4 \pi G}{k^{\prime 2}} \!\! \int \!\! \rd \bvp \!\! \int \!\! \frac{\rd \omega}{2 \pi} \frac{\rd \omegap}{2 \pi} \, \frac{1}{\eps (\bkp , \omegap)} \, \frac{\re^{- \ri (\omega + \omegap)t}}{(\omega - \bk\!\cdot\!\bv)(\omegap - \bkp\!\cdot\!\bvp)} \, \big\langle \hf_{0} (\bk , \bv) \, \hf_{0} (\bkp , \bvp) \big\rangle \bigg] ,
\label{calc_F1_F2}
\\
F_{2} (\bv) {} & = \ri \!\! \int \!\! \rd \bk \rd \bkp \, \re^{\ri(\bk+\bkp)\cdot\br}\,\hoD_{\bkp} \bigg[ \frac{(4 \pi G)^{2}}{k^{2} k^{\prime 2}} \hoD_{\bk} \big[ \oW (\bv) \big] \!\! \int \!\! \rd \bvp \rd \bvpp \, \!\! \int \!\! \frac{\rd \omega}{2 \pi} \frac{\rd \omegap}{2 \pi} \, \frac{1}{\eps (\bk , \omega) \, \eps (\bkp , \omegap)} \, \frac{\re^{- \ri (\omega + \omegap)t}}{(\omega \!-\! \bk\!\cdot\! \bv) (\omega \!-\! \bk\!\cdot\! \bvp) (\omegap \!-\! \bkp\!\cdot\!\bvpp)}
\nonumber
\\
{} & \hspace{7cm} \times \big\langle \hf_{0} (\bk , \bvp) \, \hf_{0} (\bkp  ,\bvpp) \big\rangle \bigg] .
\nonumber
\end{align}

To pursue the calculation further, we must now characterize the properties
of the correlations of the initial fluctuations in the system.
In a homogeneous system, such correlations can only depend on the positions $\br$ and $\br'$ of the two points through their difference $\br-\br'$. Moreover we assume that the particles' velocities are chosen independently from the time-averaged DF $F_\rb(\bv)$, so the correlation function must vanish if $\bv\not=\bv'$. Therefore we can write
\begin{align}
\big\langle f_0(\br,\bv)\,f_0(\brp , \bvp)\big\rangle &=\deltaD(\bv-\bvp) \, C(\br-\brp,\bv) ,
\nonumber
\\
\big\langle \hf_{0} (\bk , \bv) \, \hf_{0} (\bkp , \bvp) \big\rangle &= \deltaD (\bk + \bkp) \, \deltaD (\bvp - \bvp) \, \hC(\bk , \bv) ,
\label{eq:correl_split}
\end{align}
with $\hC(\bk,\bv)$ the Fourier transform of the correlation function $C(\br-\br',\bv)$.  Since the DF $W(\br,\bv,t)$ is real, $f_0(\br,\bv)$ is also real. Therefore $C(\br,\bv)$ is real and from its definition it is also an even function of $\br$. Thus, $\hC(\bk , \bv)$ is real and an even function of $\bk$.

We postpone to Appendix~\ref{sec:Correlation}
the explicit calculation of the function ${ \hC (\bk , \bv) }$.
Inserting Eq.~\eqref{eq:correl_split} in Eq.~\eqref{calc_F1_F2}, we get
\begin{align}
F_{1} (\bv) {} & = - \ri \!\! \int \!\! \rd \bk \, \hoD_{\bk} \bigg[ \frac{4 \pi G}{k^{2}} \, \hC (\bk , \bv) \!\! \int \!\! \frac{\rd \omega}{2 \pi} \frac{\rd \omegap}{2 \pi} \, \frac{1}{\eps (-\bk , \omegap)} \, \frac{\re^{- \ri (\omega + \omegap)t}}{(\omega - \bk\!\cdot\! \bv) (\omegap + \bk\!\cdot\!\bv)} \bigg] ,
\label{recalc_F1_F2}
\\
F_{2} (\bv) {} & = - \ri \!\! \int \!\! \rd \bk \, \hoD_{\bk} \bigg[ \frac{(4 \pi G)^{2}}{k^{4}} \hoD_{\bk} \big[ \oW (\bv) \big] \!\! \int \!\! \rd \bvp \, \hC (\bk , \bvp) \!\! \int \!\! \frac{\rd \omega}{2 \pi} \frac{\rd \omegap}{2 \pi} \, \frac{1}{\eps (\bk , \omega) \, \eps (- \bk , \omegap)} \, \frac{\re^{- \ri (\omega + \omegap) t}}{(\omega - \bk\!\cdot\! \bv)(\omega - \bk\!\cdot\! \bvp)(\omegap + \bk\!\cdot\!\bvp)} \bigg] ,
\nonumber
\end{align}
where we have used the relation ${ \hoD_{- \bk} [\oW (\bv)] = - \hoD_{\bk} [\oW (\bv)] }$.

If we assume that the system is initially stable, then all the inverse Laplace transforms present in Eq.~\eqref{recalc_F1_F2}
can be explicitly computed, as detailed in Appendix~\ref{sec:CalcLaplace}.
We get
\begin{align}
F_{1} (\bv) {} & = \ri \!\! \int \!\! \rd \bk \, \hoD_{\bk} \bigg[ \frac{4 \pi G}{k^{2}} \frac{\hC (\bk , \bv)}{|\eps (\bk , \bk\!\cdot\! \bv)|^{2}} \, \eps (\bk , \bk\!\cdot\!\bv) \bigg] ,
\nonumber
\\
F_{2} (\bv) {} & = - \ri \!\! \int\!\! \rd \bk \, \hoD_{\bk} \bigg( \frac{(4 \pi G)^{2}}{k^{4}} \hoD_{\bk} \big[ \oW (\bv) \big] \!\! \int \!\! \rd \bvp \, \frac{\hC (\bk , \bvp)}{|\eps (\bk , \bk\!\cdot\!\bvp)|^{2}} \, \bigg\{ \mP \bigg[\! \frac{1}{\bk\!\cdot\! (\bv - \bvp)} \!\bigg] + \ri \pi \deltaD [\bk\!\cdot\! (\bv - \bvp)]  \bigg\} \bigg) .
\label{rerecalc_F1_F2}
\end{align}
The dielectric function $\epsilon(\bk,\bk\!\cdot\!\bv)$ is given by equation \eqref{eq:dielectric} and the Plemelj formula \eqref{Plemelj_formula},
\begin{equation}
    \epsilon(\bk,\bk\!\cdot\!\bv)=1-\frac{4\pi G}{k^2}\mP \!\!\int\!\! \rd\bvp\frac{\hoD_{\bk}[\oW(\bvp)]}{\bk\!\cdot\!(\bv-\bvp)}+\frac{4\pi^2 G}{k^2} \, \ri \,  \!\!\int\!\!\rd \bvp\,\deltaD[\bk\cdot(\bv-\bvp)]\,\hoD_\bk\big[\oW(\bvp)\big].
\end{equation}
Since $\hoD_\bk[\oW(\bv)]$ is an odd function of $\bk$, the contribution of the real part of the dielectric function integrates to zero in the expression for ${ F_{1} (\bv) }$ (this result also follows from  the physical argument that ${ F_{1} (\bv) }$ must be real).
Similar arguments can be used to simplify the expression for ${ F_{2} (\bv) }$.
All in all, we get
\begin{align}
F_{1} (\bv) {} & = - \pi \!\! \int \!\! \rd \bk \, \hoD_{\bk} \bigg[ \frac{(4 \pi G)^{2}}{k^{4}} \!\! \int \!\! \rd \bvp \, \deltaD [\bk\!\cdot\! (\bv - \bvp)] \, \hoD_{\bk} \big[ \oW (\bvp) \big] \, \frac{\hC (\bk , \bv)}{|\eps (\bk , \bk\!\cdot\! \bv)|^{2}} \bigg] , 
\nonumber
\\
F_{2} (\bv) {} & = \pi \!\! \int \!\! \rd \bk \, \hoD_{\bk} \bigg[ \frac{(4 \pi G)^{2}}{k^{4}} \!\! \int \!\! \rd \bvp \, \deltaD [\bk\!\cdot\! (\bv - \bvp)] \, \hoD_{\bk} \big[ \oW (\bv) \big] \, \frac{\hC (\bk , \bvp)}{|\eps (\bk , \bk\!\cdot\!\bv)|^{2}} \bigg] .
\label{rererecalc_F1_F2}
\end{align}
Glancing back at Eq.~\eqref{rewrite_average},
we can finally rewrite the kinetic equation as
\begin{equation}
\frac{\p \oW}{\p t} = \pi \!\! \int \!\! \rd \bk \, \hoD_{\bk} \bigg[ \frac{(4 \pi G)^{2}}{k^{4}} \!\! \int \!\! \rd \bvp \, \frac{\deltaD [\bk\!\cdot\! (\bv - \bvp)]}{|\eps (\bk , \bk\!\cdot\!\bv)|^{2}} \, \bigg( \hoD_{\bk} \big[ \oW (\bv) \big] \, \hC (\bk , \bvp) - \hoD_{\bk} \big[ \oW (\bvp) \big] \, \hC (\bk , \bv) \bigg) \bigg] .
\label{FuzzyKadomtsev}
\end{equation}
Equation~\eqref{FuzzyKadomtsev} is equivalent to Eq.~{(9)}
in~\citet{Kadomtsev+1970} except that there the
gravitational interactions are replaced with Coulomb interactions,
paying careful attention to the change of sign of the attraction,
while the finite-difference operators
are replaced with their classical limits, as in Eq.~\eqref{classical_derivatives}.
In order to finalize the calculation, it now only remains to compute explicitly
the autocorrelation function, ${ \hC (\bk , \bv) }$,
as defined in Eq.~\eqref{eq:correl_split}.

We compute this correlation function in Appendix~\ref{sec:Correlation},
\begin{equation}
\hC (\bk , \bv) = \frac{1}{(2 \pi)^{3}} \bigg[ \mb + \frac{h^{3}}{\mb^{3}} \, \oW (\bv) \bigg] \, \oW (\bv) .
\label{res_C}
\end{equation}
As a consequence, Eq.~\eqref{FuzzyKadomtsev} becomes
\begin{align}
  \frac{\p \oW (\bv)}{\p t}
  =
    2 G^2 \!\!\int\!\! \rd \bk \,
    \!\!\int\!\! \rd \bvp \, \hoD_\bk
    \bigg\{
   {} &
   \frac{\deltaD [\bk \!\cdot\! (\bv - \bvp)]}{k^4 |\eps(\bk , \bk \!\cdot\! \bv)|^2} 
    \hoD_\bk \big[ \oW (\bv) \big] \oW (\bvp)
    \bigg[ \mb +
   \frac{h^{3}}{\mb^{3}} \oW (\bvp) \bigg]
   \nonumber
   \\
    - {} &
    \frac{\deltaD [\bk \!\cdot\! (\bv - \bvp)]}{k^4 |\eps(\bk , \bk \!\cdot\! \bv)|^2} 
    \hoD_\bk \big[ \oW (\bvp) \big]
    \oW (\bv)
    \bigg[ \mb +
    \frac{h^{3}}{\mb^{3}} \oW (\bv) \bigg]
    \bigg\} .
  \label{eq:BL_II_almost}
\end{align}
The FP equation is based on the approximation that deflections are weak or that the momentum change due to gravitational scattering is small (see discussion preceding Eq.\ \ref{eq:calc_dfdt}). In the present context, this corresponds to the approximation that $\bk$ is small compared to the characteristic scale of changes in $\oW(\bv)$. In this case we can replace the discrete derivatives in $\hoD_\bk[\oW(\bv)]$ with their continuous analogs, as in the first of  Eqs.~\eqref{classical_derivatives}. 
Equation~\eqref{eq:BL_II_almost} then becomes
\begin{align}
\label{eq:BL_II}
  \frac{\p \oW (\bv)}{\p t} = {} &
    2 G^2 \!\!\int\!\! \rd \bk \,
    \bk \!\cdot\! \frac{\p }{\p \bv}
    \!\!\int\!\! \rd \bvp \, 
    \frac{\deltaD [\bk \!\cdot\! (\bv - \bvp)]}{k^4{|\eps(\bk , \bk \!\cdot\! \bv)|}^2} 
    \nonumber
    \\
  {} &
    \qquad\qquad
    \times
    \bigg\{
    \bk \!\cdot\! \frac{\p \oW (\bv) }{\p \bv}
    \oW (\bvp)
    \bigg[ \mb +
    \frac{h^{3}}{\mb^{3}} \oW (\bvp) \bigg]
    -
    \bk \!\cdot\!
    \frac{\p \oW (\bvp)}{\p \bvp}
    \oW (\bv)
    \bigg[ \mb +
    \frac{h^{3}}{\mb^{3}} \oW (\bv) \bigg]
    \bigg\} .
\end{align}
Equation~\eqref{eq:BL_II} is the main result of this section,
a \BL\ type kinetic equation
describing the self-consistent relaxation of a homogeneous \FDM\ halo.
In particular, we note that this kinetic equation involves the dielectric function,
${ 1/|\eps|^{2} }$, which describes how the fluctuations are dressed by collective effects.

In Section~\ref{sec:Stability} we show that this system is unstable for perturbations with wavenumber smaller than
the effective Jeans scale ${ \sim \min\{ \kJ , \tkJ \} }$, where $\kJ$ is the classical Jeans wavenumber and $\tkJ$ its quantum analog.
Assuming that the system is much smaller than the effective Jeans allows us to neglect collective effects and set ${ 1/|\eps|^{2} \to 1 }$.
Equation~\eqref{eq:BL_II} finally becomes
\begin{align}
  \label{eq:BL_ms_II}
  \frac{\p \oW(\bv)}{\p t}
  = {} &
    2G^2\ln\Lambda
    \frac{\p }{\p v_i}
    \!\!\int\!\! \rd \bvp \, 
    u_{ij}(\bv-\bvp)
    \nonumber
    \\
  {} &
   \times \bigg\{
    \frac{\p \oW(\bv)}{\p v_j}
    \oW(\bvp)
    \bigg[ \mb +
    \frac{h^{3}}{\mb^{3}} \oW(\bvp) \bigg]
    -
    \frac{\p \oW (\bvp)}{\p \vp_{j}}
    \oW(\bv)
    \bigg[ \mb +
    \frac{h^{3}}{\mb^{3}} \oW(\bv) \bigg]
    \bigg\}
\end{align}
where $u_{ij}(\bv)$ is defined in Eq.~\eqref{eq:uij_I}.
We have therefore reached our final result,
as we have recovered the kinetic equation \eqref{eq:landau_fdm} describing the self-consistent relaxation of a homogeneous \FDM\ halo.

\bigskip
\section{Linear Stability}
\label{sec:Stability}

As for many physical systems, the dielectric function \eqref{eq:dielectric} is central to understanding the dynamical behavior of an FDM halo. To illustrate the utility of this function we explore the stability properties of the halo. 
Throughout this section, we will assume for simplicity
that the unperturbed \DF\ is isotropic, ${ \Fb (\bv) = \Fb (v) }$.

In the classical or particle limit, an infinite homogeneous system is susceptible to an instability characterized by the classical Jeans wavenumber $\kJ$ \citep{Jeans1902,Binney+2008}. If the DF is Maxwellian,
\begin{equation}
\Fb (\bv) = \frac{\rho_{0}}{(2 \pi \sigma^{2})^{3/2}} \, \re^{- |\bv|^{2}/(2 \sigma^{2})} ,
\label{Maxwellian_DF}
\end{equation}
then 
\begin{equation}
\kJ = \frac{(4 \pi G \rho_{0})^{1/2}}{\sigma} ,
\label{def_kJ}
\end{equation}
and perturbations with ${ k \leq \kJ }$ are unstable,
while ones with ${ k > \kJ }$ are stable.

In contrast, a halo composed of waves rather than particles that has zero velocity dispersion (i.e., the unperturbed wavefunction $\psi(\br,t)=\mbox{cst.}$) is unstable to perturbations with wavenumber $k<\tkJ$ where the quantum Jeans wavenumber is 
\citep[see, e.g.\@,][]{Khlopov+1985,Bianchi+1990,Hu+2000,Chavanis2011}
\begin{equation}
\tkJ = 2 \bigg( \frac{\pi G \rho_{0} \mb^{2}}{\hbar^{2}} \bigg)^{1/4}.
\label{def_tildekJ}
\end{equation}

In an FDM halo with non-zero velocity dispersion,
the effective Jeans wavenumber can be determined from the dielectric function \eqref{eq:dielectric}. For an isotropic DF,
\begin{equation}
     \label{eq:dielectric_iso}
     \eps (k, \omega) = 1 - \frac{4\pi G\mb}{k^2\hbar} \!\! \int \!\! \frac{\rd u}{\omega-ku} \, \bigg[ \Fpara \bigg( u + \frac{\hbar k}{2 \mb} \bigg) - \Fpara \bigg( u - \frac{\hbar k}{2 \mb} \bigg) \bigg],
\end{equation}
with ${ \Fpara (u) \!=\! \!\int\! \rd v_1 \rd v_2 \, \Fb [(v_1^2+v_2^2+u^2)^{1/2}] }$ and $v_1$ and $v_2$ are the velocities along the two axes perpendicular to $\bk$. 
The system is linearly unstable if there exists a frequency $\omega$
in the upper half of the complex plane and a real wavenumber $k$ such that ${ \eps (k, \omega) = 0 }$. In order to investigate the system's stability,
we place ourselves at the limit of marginal
stability, i.e.,\ we assume that ${ \ImPart (\omega) \to 0 }$.
In that case, we can use the Plemelj formula
from Eq.~\eqref{Plemelj_formula}
to rewrite Eq.~\eqref{eq:dielectric_iso} as
\begin{align}
     \label{eq:dielectric_iso_p}
     \eps (k, \omega) = {} &1 - \frac{4\pi G \mb}{k^2\hbar}\mathcal{P} \!\! \int \!\! \frac{\rd u}{\omega - ku} \, \bigg[ \Fpara \bigg( u + \frac{\hbar k}{2 \mb} \bigg) - \Fpara \bigg( u - \frac{\hbar k}{2 \mb} \bigg) \bigg]
     \nonumber
     \\
     {} & + 
 \ri\frac{4\pi^2 G\mb}{k^3\hbar}
     \bigg[ \Fpara \bigg( \frac{\omega}{k} + \frac{\hbar k}{2 \mb} \bigg)
     - \Fpara \bigg( \frac{\omega}{k} -\frac{\hbar k}{2 \mb} \bigg) \bigg].
\end{align}
In order to have ${ \eps (k , \omega) = 0 }$,
both the real part and the imaginary part of Eq.~\eqref{eq:dielectric_iso_p}
must vanish. Let us assume for simplicity that the \DF\ $\Fb(v)$
is a monotonic decreasing function of $v=|\bv|$ (i.e., we ignore two-stream instabilities).
Then it is straightforward to show that $\Fpara (u)$ is an even function of $u$,  monotonic decreasing for $u>0$ and  monotonic increasing for $u<0$.  As a consequence, 
the second term from Eq.~\eqref{eq:dielectric_iso_p}
vanishes if and only if ${ k = 0 }$ or ${ \omega = 0 }$.
Because that first possibility is not of physical interest,
we may then assume that ${ \omega = 0 }$
when investigating the system's marginal stability.

For simplicity, let us now assume that the system's unperturbed \DF\ is Maxwellian
(see Eq.~\eqref{Maxwellian_DF}). In that case, we can rewrite
the dielectric function from Eq.~\eqref{eq:dielectric_iso} as
\begin{align}
\eps (k , \omega) {} & = 1 - \frac{4 \pi G \mb \rho_{0}}{\sqrt{2 \pi} \, k^{2} \sigma \hbar} \! \int \!\! \frac{\rd v}{\omega - k v} \, \bigg[ \re^{- [v + \hbar k / (2 \mb)]^{2} / (2 \sigma^{2})} - \re^{- [v - \hbar k / (2 \mb)]^{2} / (2 \sigma^{2})} \bigg]
\nonumber
\\
{} & = 1 - \bigg( \frac{\kJ}{k} \bigg)^{3} \, \frac{1}{2 \sqrt{\pi} \eta} \!\! \int \!\! \frac{\rd x}{\varpi - x} \, \bigg[ \re^{- [x + k \eta / (2 \kJ)]^{2}} - \re^{- [x - k \eta / (2 \kJ)]^{2}} \bigg]
\nonumber
\\
{} & = 1 - \bigg( \frac{\kJ}{k} \bigg)^{3} \frac{1}{2 \eta} \!\bigg[ Z \bigg( \varpi - \frac{k \eta}{2 \kJ} \bigg) - Z \bigg( \varpi + \frac{k \eta}{2 \kJ} \bigg) \bigg]
\label{dielectric_Maxwellian}
\end{align}
In that expression, we introduced
the rescaled frequency ${ \varpi = \omega/(\sqrt{2} k \sigma) }$,
as well as the dimensionless ratio
\begin{align}
\eta {} & = \sqrt{2} \, \bigg( \frac{\kJ}{\tkJ} \bigg)^{2} = \frac{\sqrt{2 \pi G \rho_{0}} \hbar}{\sigma^{2} \mb}
\nonumber
\\
{} & \simeq 0.0315 \, \bigg( \frac{\mb}{10^{-22} \eV} \bigg)^{-1} \bigg( \frac{\rho_{0}}{0.01 \Msun \pc^{-3}} \bigg)^{1/2} \bigg( \frac{\sigma}{100 \kms} \bigg)^{-2} .
\label{def_eta}
\end{align}
We note that ${ \eta \ll 1 }$ corresponds
to the classical limit,
while ${ \eta \gg 1 }$ is associated with the quantum limit.
In the last line of Eq.~\eqref{dielectric_Maxwellian},
we introduced the plasma dispersion function~\citep[see, e.g.\@,][]{Fried+1961},
defined as
\begin{align}
Z (\varpi) {} & = \frac{1}{\sqrt{\pi}} \!\! \int_{- \infty}^{\infty} \!\!\!\! \rd s \, \frac{\re^{-s^{2}}}{s - \varpi}
=
\begin{cases}
\displaystyle \! \int_{- \infty}^{\infty} \!\!\!\! \rd s \, \frac{\re^{ - s^{2}}}{s - \varpi} {} & \text{ if } \quad \ImPart (\varpi) > 0 ,
\\
\displaystyle \mP \!\! \int_{- \infty}^{ \infty} \!\!\!\! \rd s \, \frac{\re^{-s^{2}}}{s - \varpi} + \pi\ri \re^{-\varpi^{2}} {} & \text{ if } \quad \ImPart (\varpi) = 0 ,
\\
\displaystyle \! \int_{- \infty}^{\infty} \!\!\!\! \rd s \, \frac{\re^{-s^{2}}}{s - \varpi} + 2 \pi\ri \re^{- \varpi^{2}} {} & \text{ if } \quad \ImPart (\varpi) < 0 .
\end{cases}
\label{def_Z}
\end{align}

\begin{figure}[ht]
\epsscale{0.7}
\plotone{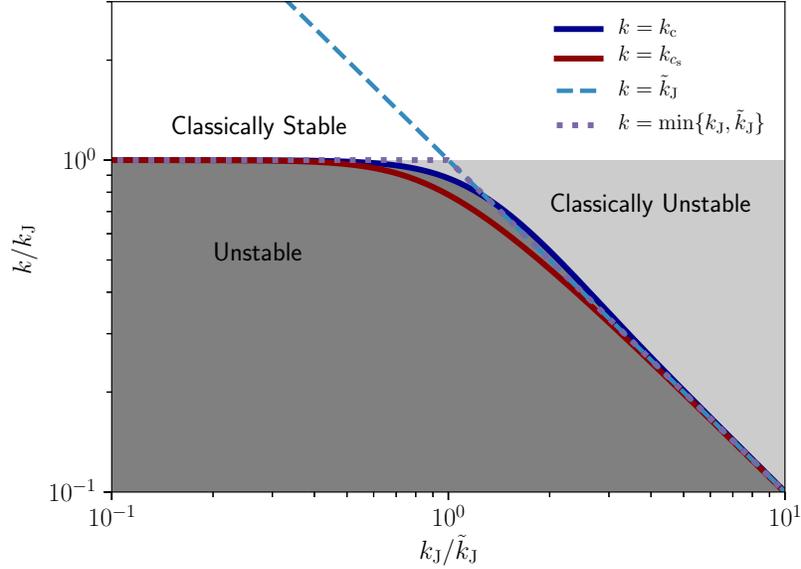}
\caption{Linear stability of a Maxwellian DF. The critical wavenumber $\kc$ that separates stable and unstable perturbations is shown 
as a function of the ratio between the classical Jeans wavenumber $\kJ$ and the quantum Jeans wavenumber
  $\tkJ$ (blue line).
  Perturbations with wavenumber  ${ k > \kc }$ are
  stable and perturbations with wavenumber ${ k < \kc }$ are unstable.
  This critical wavenumber $\kc$ can be approximated by ${ \min \{ \kJ , \tkJ \} }$ (dotted line).
  The critical wavenumber $\kcs$ for a fluid system (see Eq.~\ref{def_kcs} and \citealt{Chavanis2011})
  is also shown. 
\label{fig:k0}}
\end{figure}

Following Eq.~\eqref{dielectric_Maxwellian}, the requirement of 
marginal stability at ${ \varpi = 0 }$
now leads to the implicit relation
\begin{equation}
\kc^{3} = \frac{\kJ^{3}}{2 \eta} \bigg[ Z \bigg( -\frac{\kc \eta}{2 \kJ} \bigg) - Z \bigg( \frac{\kc \eta}{2 \kJ} \bigg) \bigg] ,
\label{kc_Maxwell}
\end{equation}
with $\kc$ the critical wavenumber at marginal stability.
For ${ x \in \mathbb{R} }$, we can use the two expansions~\citep{Fried+1961}
\begin{align}
Z (x) \simeq
\begin{cases}
\displaystyle \ri \sqrt{\pi} \re^{-x^{2}} - 2 x {} & \text{ for } \;\; |x| \ll 1 ,
\\
\displaystyle \ri \sqrt{\pi} \re^{-x^{2}} - \frac{1}{x} {} & \text{ for } \;\; |x| \gg 1 .
\end{cases}
\label{relations_Z}
\end{align}
Then we can approximate the r.h.s.\ of Eq.~\eqref{kc_Maxwell} as
\begin{equation}
\kc \simeq
\begin{cases}
\displaystyle \kJ & \text{ for } \;\; \eta \ll 1 , 
\\
\displaystyle \tkJ & \text{ for } \;\; \eta \gg 1 .
\end{cases}
\label{expansion_kc_Maxwell}
\end{equation}
We therefore recover the known result that in the limit ${ \eta \ll 1 }$ (${ \kJ \ll \tkJ }$), the system's stability is determined by classical physics and the critical wavenumber is $\kJ$, while in the limit ${ \eta \gg 1 }$ (${ \kJ \gg \tkJ  }$), stability is dominated by quantum effects and the critical wavenumber is $\tkJ$.
This behaviour is illustrated in Fig.~\ref{fig:k0},
where we show the critical wavenumber for a Maxwellian DF as determined from Eq.~\eqref{kc_Maxwell}, 
as a function of the ratio ${ \kJ / \tkJ }$.

\cite{Chavanis2011} derived a simple dispersion relation
by assuming that the fuzzy halo is a fluid with
a sound speed $\cs$.
Assuming that this parameter is a proxy for the halo velocity dispersion,
i.e.,\ ${ \cs = \sigma }$,
Eq.~{(138)} of~\cite{Chavanis2011} gives the simple dispersion relation
\begin{equation}
\kcs^{2} = \frac{\kJ^{2}}{\eta^{2}} \bigg[ \sqrt{1 + 2 \eta^{2}} - 1 \bigg] ,
\label{def_kcs}
\end{equation}
so that, in this model,
perturbations with ${ k > \kcs }$ are stable,
while ones with ${ k < \kcs }$ are unstable.
The prediction of Eq.~\eqref{def_kcs} is also illustrated
in Fig.~\ref{fig:k0}; it correctly
recovers the transition
between the classical and quantum regimes
as one varies the ratio ${ \kJ / \tkJ }$.

\begin{figure}[ht]
\epsscale{0.7}
\plotone{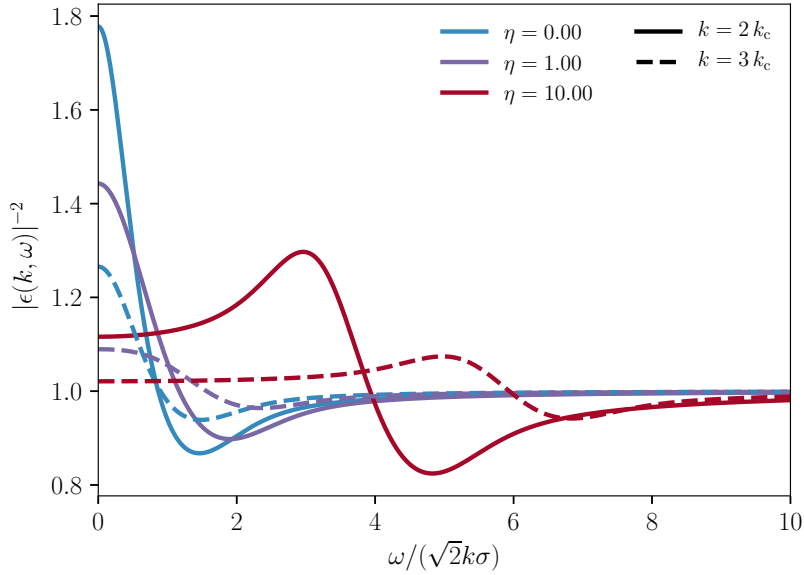}
\caption{
Illustration of the self-gravitating dressing of perturbations by collective effects,
as measured by the dielectric function ${ 1/|\eps (k , \omega)|^{2} }$,
for different values of the stability parameter ${ k / \kc }$,
and the quantum parameter $\eta$.
\label{fig:eps}}
\end{figure}

In Fig.~\ref{fig:eps}, we illustrate the importance of the dressing of relaxation by collective effects,
through the factor ${ 1/|\eps (k , \omega)|^{2} }$ as appearing in Eq.~\eqref{eq:BL_II}.
 Here the dielectric function $\epsilon(k,\omega)$ is given by 
Eq.~\eqref{dielectric_Maxwellian}.
We note that this dressing becomes negligible (i.e.,\ ${ \eps (k , \omega) \to 1 }$)
as the system becomes more stable (i.e.,\ ${ k \gg \kc }$).
Moreover, for a given value of ${ k / \kc }$,
as quantum effects become more important (i.e.,\ ${ \eta \gg 1 }$) 
collective effects become negligible, $\eps(k,\omega) \sim \eta^{-3}$.

\bigskip

\section{Numerical application}
\label{sec:Application}

In this final section we present some numerical simulations of \FDM\ halos,
and compare them with the predictions from the kinetic theory derived above.
In order to investigate the long-term relaxation
of the system, we also present direct time integrations
of the diffusion equation itself,
in particular highlighting the unavoidable formation of the central soliton
for cold enough initial conditions.

We use numerical methods similar to those of~\cite{Levkov+2018}.
Importantly, in order to evade the Jeans instability,
we assume ${ G < 0 }$,
which ensures that the system is stable and does not affect kinetic equations such as \eqref{eq:landau_fdm} as they only contain $G^{2}$.
We consider a three-dimensional box of length $L$
that is discretized in ${ K^{3} }$ cells.
Each location on the grid is characterized
by a position ${ \br = \Delta_{r} \bn }$,
with ${ \Delta_{r} = L / K }$
and ${ \bn \in \{ 0, ... , K \!-\!1 \}^{3} }$.
At each grid location, we track the local value of the wavefunction ${ \psi_{\br} (t) }$ as well as the gravitational potential ${ \Phi_{\br} (t) }$.

The initial conditions are set so that the wavefunction
approximates a uniform density Maxwellian \DF\@,
as defined in Eq.~\eqref{Maxwellian_DF}.
To do so, we naturally perform the discrete Fourier expansion
\begin{equation}
\hpsi_{\bk} (t) = \frac{\Delta_{r}^{3}}{(2 \pi)^{3}} \sum_{\br} \psi_{\br} (t) \, \re^{-\ri \bk \cdot \br} ;
\quad
\psi_{\br} (t) = \Delta_{k}^{3} \sum_{\bk} \hpsi_{\bk} (t) \, \re^{\ri \bk \cdot \br} , 
\label{discrete_Fourier}
\end{equation}
where we introduced ${ \bk = \Delta_{k}\bn}$,
with ${ \Delta_{k} = 2 \pi / L }$ and $\bn\in {\{-K/2,\dots,-1,1,\dots,K/2\}}^3$ .
Each of the Fourier wavenumbers is then initialized with
\begin{equation}
\hpsi_{\bk} (0) = \sqrt{f_{k} (k)} \, \re^{\ri \phi_{\bk}} ;
\quad
f_{k} (k) = \frac{\hbar^{3}}{\Delta_{k}^{3} \mb^{3}} \, \Fb \big( v = \hbar k/\mb\big) ,
\label{init_hpsi}
\end{equation}
where ${ \phi_{\bk} }$ is a random phase uniformly
distributed in ${ [0 , 2 \pi] }$
and uncorrelated on the $\bk$-grid, that is 
${ \langle \phi_{\bk} \, \phi_{\bkp} \rangle \propto \delta_{\bk\bkp} }$.
Once the wavefunction is known, we can compute its associated density ${ | \psi_{\br} |^{2} }$.
This is subsequently used in the Poisson equation~\eqref{eq:schr-poisson}
to estimate the potential $\Phi_{\br}$.
The calculation of the potential is performed in Fourier space,
using a FFT (i.e.,\ assuming periodic boundary conditions),
and further accelerated by GPU.

Once ${ \psi_{\br} (t) }$ and ${ \Phi_{\br} (t) }$ are known,
we may proceed with the forward integration in time of the SP equations. This is performed through appropriate sequences of kick and drift operators, given by
\begin{align}
\text{Drift: } {} & \quad
\hpsi_{\bk} \to \hpsi_{\bk} \, \re^{- \ri \hbar \Delta t |\bk|^{2} / (2 \mb)} ,
\nonumber
\\
\text{Kick: } & {} \quad
\psi_{\br} \to \psi_{\br} \, \re^{- \ri \Delta t \mb \Phi_{\br}/\hbar} .
\label{drift_operators}
\end{align}
The timestep ${ \Delta t }$ and order of the integrator have to be picked carefully.
In practice, we used a sixth-order explicit symplectic
integrator~\citep{Yoshida1990}.
Once we are able to perform numerical simulations of the system,
we measure the value of ${ \p \Fb (v = \hbar k / \mb) / \p t }$
by fitting the function ${ t \to \langle | \hpsi_{\bk} (t) |^{2} \rangle \Delta_{k}^{3} \mb^{3}  / \hbar^{3} }$
with a linear function of $t$,
and performing an ensemble average over $10^{3}$ realisations
with different initial conditions.

Having integrated the \FDM\ dynamics,
we may now compare the results with the prediction
from kinetic theory.
In the limit where collective effects can be neglected,
the system's relaxation is described by Eq.~\eqref{eq:landau_fdm}.
In particular, for an isotropic system
with a Maxwellian \DF\@, as in Eq.~\eqref{Maxwellian_DF}, We note that 
the flux generated by the classical diffusion coefficients $\Dc_{i}$ and $\Dc_{ij}$ (see Eq.~\ref{eq:D_cb}) vanishes exactly,
so the only surviving contributions
are from the wave diffusion coefficients $\Db_i$ and $\Db_{ij}$. With this simplification we can obtain from Eq.~\eqref{eq:landau_fdm_iso}
the diffusion flux at ${ t = 0 }$ through
\begin{equation}
\frac{\p \Fb (v)}{\p t} \bigg|_{t = 0} = \frac{(2 \pi)^{1/2} G^{2} \hbar^{3} \rho_{0}^{3} \ln \Lambda}{2 \mb^{3} \sigma^{9}} \bigg[ 18 \, \re^{- 3 v^{2} / (2 \sigma^{2})} + \pi^{1/2} \frac{\sigma}{v} \, \re^{- v^{2} / (2 \sigma^{2})} \erf (v / \sigma) - 2^{7/2} \pi^{1/2} \frac{\sigma}{v} \, \re^{-v^{2} / \sigma^{2}} \erf [v / (2^{1/2} \sigma)] \bigg] ,
\label{eq:dfdt_maxwell}
\end{equation}
an exact quantitative expression of the approximate relaxation time \eqref{eq:intro_II}.

\begin{figure}[htbp!]
\epsscale{0.7}
\plotone{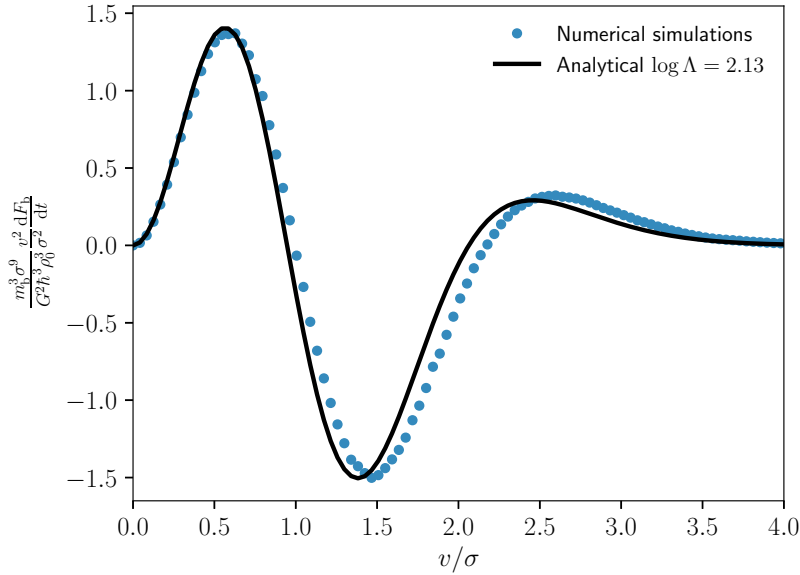}
\caption{The initial rate of change of a Maxwellian \DF\@,
as measured in the numerical simulations is in a good
  agreement with the analytical prediction
  from Eq.~\eqref{eq:dfdt_maxwell} with
  ${ \ln \Lambda \simeq 2.1 }$.
  For our numerical experiments,
 we used a box of size ${ L \!=\! 150 \!\times\! \lbarsig }$
with ${ \lbarsig \!=\! \hbar/(\mb\sigma) }$,
discretized in ${ 256^{3} }$ cells.
We picked the initial density to be ${ \rho_{0} \!=\! 50 \!\times\! \mb \lbarsig }$
and fixed the (negative) gravitational constant to ${ G = -0.06 \!\times\! \sigma\hbar /\mb }$.
\label{fig:dfdt}}
\end{figure}

In Fig.~\ref{fig:dfdt},
we compare the numerical simulations
with the prediction from Eq.~\eqref{eq:dfdt_maxwell},
using an estimate ${ \ln\Lambda \simeq 2.1 }$.
As illustrated in this figure,
we recover a good agreement between the numerical simulations
and the kinetic prediction.

In addition to integrating the \SP\ equations directly,
we investigated the evolution of the \DF\ itself
by directly integrating forward in time
the isotropic Landau equation~\eqref{eq:landau_fdm_iso}.
To do so, we divide the interval ${ 0 \leq v \leq \vmax }$
onto a regular grid.
At each of the grid locations,
the isotropic integrals from Eq.~\eqref{eq:landau_fdm_iso}
are computed using explicit second-order integration rules.
Once the evolution rate, ${ \p \Fb (v) / \p t }$, has been determined
on the velocity grid,
we integrate it forward in time using a first-order explicit Euler method
with a timestep given by ${ \Delta t = 10^{-3} \!\times\! \min_{v} [\Fb (v) / (\p \Fb (v) / \p t)] }$.

\begin{figure}[htbp!]
\epsscale{0.55}
\plotone{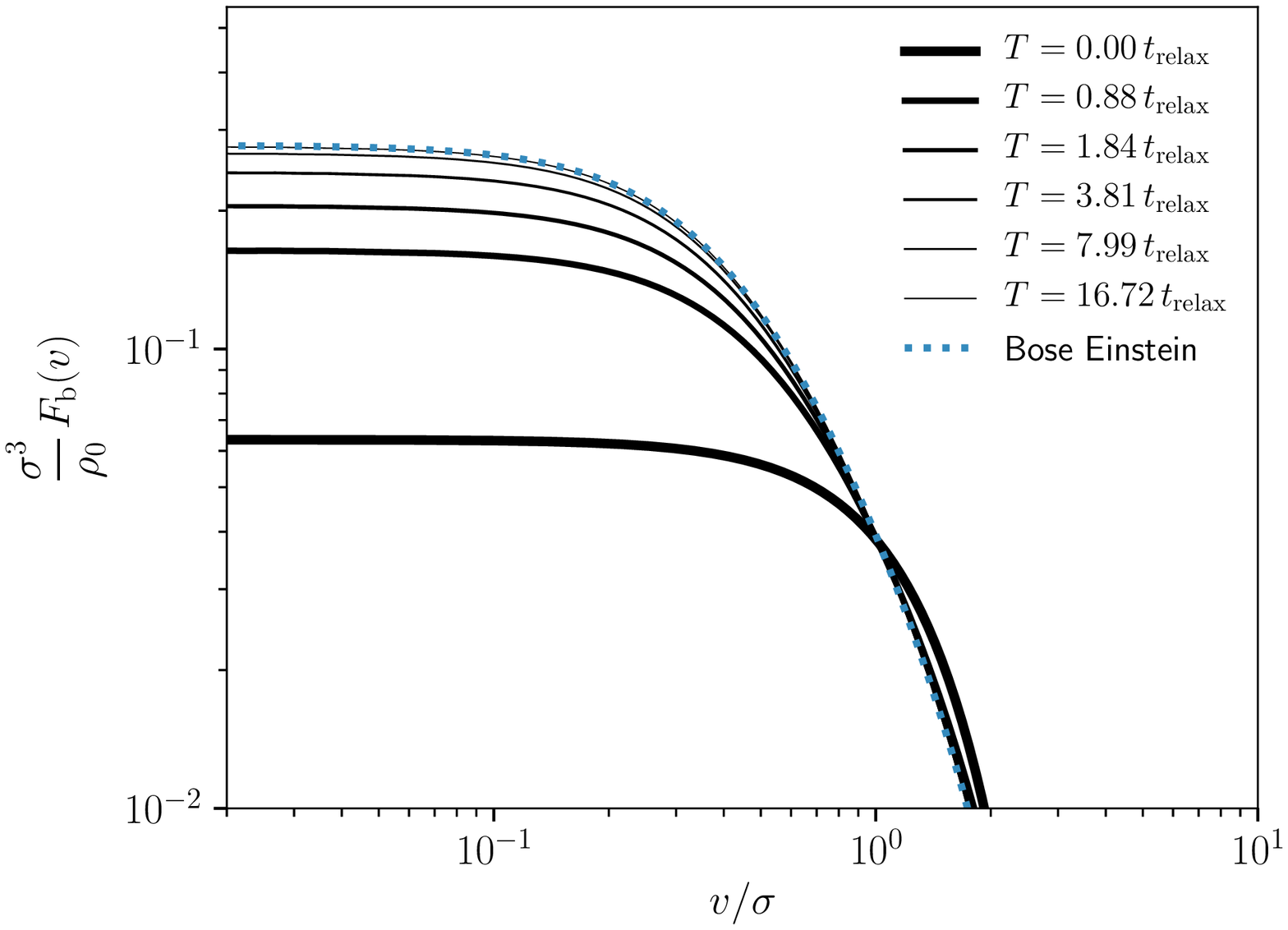}
\plotone{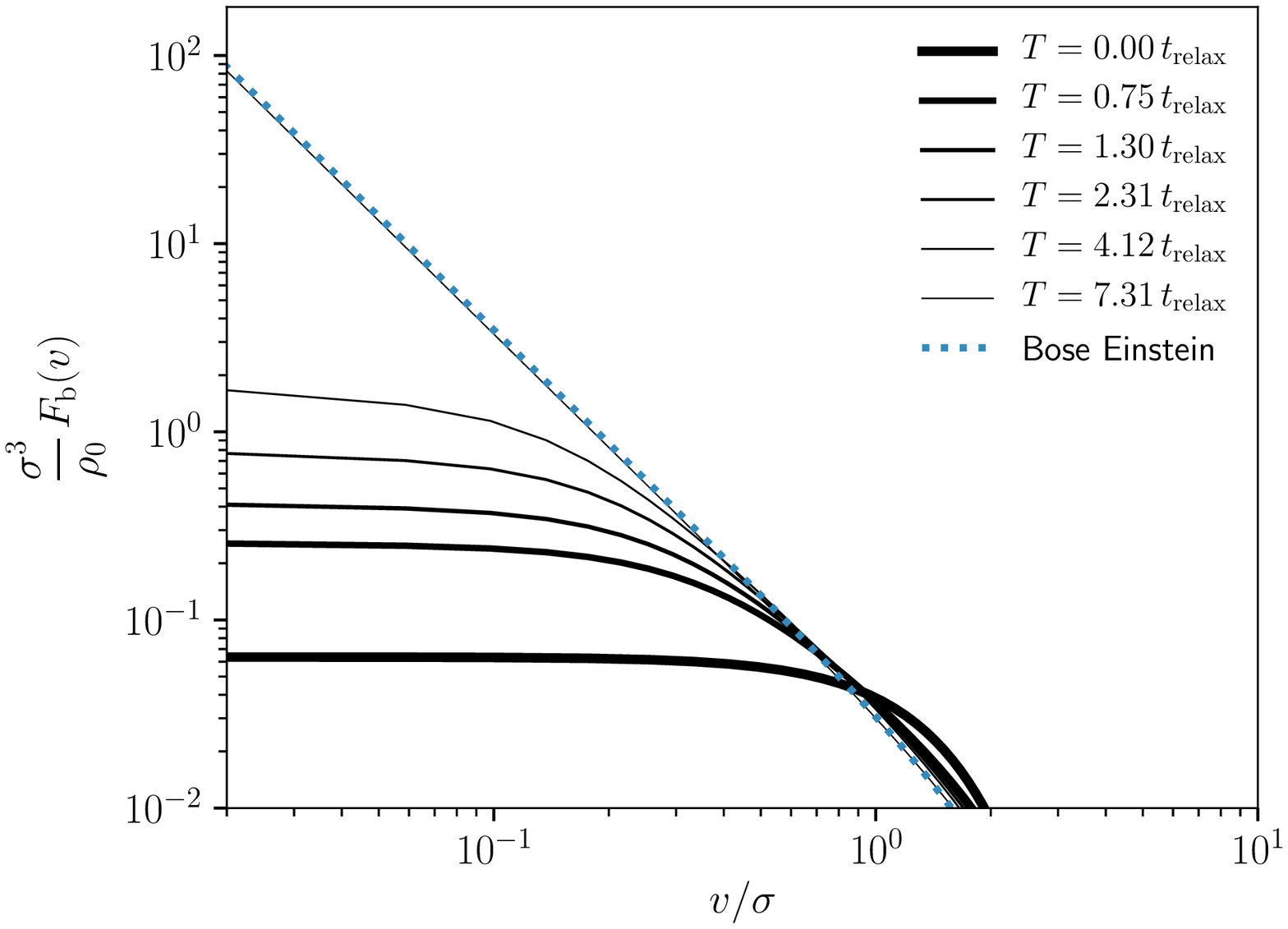}
\caption{Relaxation of a Maxwellian \DF\ with ${ \sigma = 1.3 \sigmac }$ (left panel)
and ${ \sigma = \sigmac }$ (right panel).
Here, the integration of Eq.~\eqref{eq:landau_fdm_iso}
was performed on a linear grid in $v$ with $2^{11}$ cells and ${ \vmax = 10 \!\times\! \sigma }$.
Time has been rescaled as a function of $\trelax$,
as defined in Eq.~\eqref{eq:trlx}.
\label{fig:relax_to_be}}
\end{figure}

Using that method, we show in Fig.~\ref{fig:relax_to_be}
that a Maxwellian \DF\ with ${ \sigma \simeq 1.3 \sigmac }$
(see Eq.~\eqref{eq:rho-sigma-minimum})
relaxes to a Bose-Einstein steady state
in a few relaxation times (as defined in Eq.~\eqref{eq:trlx}). 
In that same figure, we also recover that when ${ \sigma = \sigmac }$,
the \DF\ develops a ${ \Fb \propto 1/v^2 }$ cusp at small $v$ on finite time,
so that the system is almost on the verge of forming the central soliton.

\bigskip

\section{Conclusions}
\label{sec:Conclusion}

In this paper, we investigated the self-consistent relaxation of an \FDM\ halo
driven by the unavoidable and undamped quantum fluctuations
that it must sustain.
The main result was presented in Eq.~\eqref{eq:landau_fdm},
which is the appropriate generalisation of the classical Fokker--Planck (\FP\@) equation to the quantum case.
We showed how this kinetic equation can be derived either from the heuristic Boltzmann--Nordheim--Uehling--Uhlenbeck (\BNUU\@) equation
(Section~\ref{sec:DerivationBNUU}),
or from the quasi-linear perturbation of the Schr\"{o}dinger--Poisson (\SP\@) system
(Section~\ref{sec:DerivationKlim}).
We showed in particular how the diffusion can be accelerated through collective effects that can dress the perturbations. The strength of the collective effects is encapsulated in the dielectric function (Eq.\ \ref{eq:dielectric}), 
as illustrated in Eq.~\eqref{eq:BL_II} with a Balescu--Lenard (\BL\@) type kinetic equation.
We subsequently described in Section~\ref{sec:Stability}
the linear stability of a homogeneous \FDM\ halo,
making clear the connection between the classical and quantum limits.
Finally, we illustrated some of these results in Section~\ref{sec:Application}
using tailored numerical simulations.

Of course, the present paper is only a first step towards
a description of the evolution of an \FDM\ halo.
First, it is important to extend the present derivation to inhomogeneous \FDM\ halos. So long as the typical de Broglie wavelength $\lbarsig$ is small compared to the size of the halo, a good first approximation would be to treat the FDM diffusion coefficients \eqref{Di_Rosenbluth} and \eqref{Dij_Rosenbluth} as local diffusion coefficients in an inhomogeneous FP equation analogous to \eqref{eq:fp_fdm}.
For more accurate analyses, one would benefit in particular
from the recent progress in the context of classical
self-gravitating systems that recently led to the derivation of the inhomogeneous \BL\ equation~\citep{Heyvaerts2010,Chavanis2012}.
As highlighted in Figure~\ref{fig:relax_to_be},
\FDM\ halos with a sufficiently small velocity dispersion---which includes all galaxy velocity dispersions if the particle mass $\mb\ll 1\eV$---unavoidably relax to a Bose-Einstein condensate.
Describing the formation and growth of that condensate is essential for understanding the long-term fate of FDM halos.
Finally, should \FDM\ prove to be a viable alternative to classical \CDM\@,
the present kinetic theory needs to be implemented
in a cosmological context~\citep[see, e.g.\@,][]{Amin2019}.

\acknowledgments
BB is supported by the Martin A. and Helen Chooljian Membership at the Institute for Advanced Study.
JBF acknowledges support from Program HST-HF2-51374
provided by NASA through a grant from the Space Telescope Science Institute, which is operated by the Association of Universities for Research in Astronomy, Incorporated, under NASA contract NAS5--26555.
ST is supported in part by NSERC.
This work is partially supported by grant Segal ANR-19-CE31-0017 of the French Agence Nationale de la Recherche.

\appendix

\bigskip

\section{Computing the inverse Laplace transforms}
\label{sec:CalcLaplace}

In this section, we compute explicitly the Laplace transforms from Eq.~\eqref{recalc_F1_F2}.
The approach we follow is very similar to the one from~\cite{Schekochihin2017}.

Let us first start with the inverse Laplace transforms appearing
in the expression for ${ F_{1} (\bv) }$ in Eq.~\eqref{recalc_F1_F2}.
We want to evaluate
\begin{equation}
I (\bk , \bv ) = \!\! \int \! \frac{\rd \omega}{2 \pi} \frac{\rd \omegap}{2 \pi} \, \frac{1}{\eps (-\bk , \omegap)} \, \frac{\re^{- \ri (\omega + \omegap)t}}{(\omega - \bk\!\cdot\bv) (\omegap + \bk\!\cdot\! \bv)} .
\label{intLaplace_F1}
\end{equation}
Recall that each integral is along a Bromwich contour $\mB$, a horizontal contour that passes above all the poles of the integrands.
We assume that the system is linearly stable,
so the function ${ \omega \to 1 / \eps (\bk  , \omega) }$ has no poles
in the upper half of the complex plane. We carry out the integrals by lowering the integration contours to very negative imaginary values, so that ${ \re^{- \ri (\omega + \omegap)t} }$ vanishes. Assuming that $t$ is large enough for the transients associated with the system's damped modes (i.e., the contributions from the poles of the dielectric functions) 
to be negligible, only the contributions from the poles on the real axis 
${ \omega = \bk \!\cdot\! \bv }$ and ${ \omegap = - \bk\!\cdot\!\bv }$ remain.
Paying careful attention to the direction of integration, we note that these poles each contribute ${ - 2 \pi \ri }$ times the associated residue.
Using these arguments, we obtain
\begin{equation}
I (\bk , \bv) = - \frac{\eps (\bk , \bk\!\cdot\! \bv)}{|\eps (\bk , \bk\!\cdot\!\bv)|^{2}} .
\label{calc_Laplace_F1}
\end{equation}
Here we have used the symmetry relation
\begin{equation}
\eps (- \bk , - \omegaR) = \eps^{*} (\bk , \omegaR) ,
\label{symm_eps}
\end{equation}
for ${ \omegaR \in \mathbb{R} }$, which directly follows from Eq.~\eqref{eq:dielectric}.

The second integral to compute appears in the expression for ${ F_{2} (\bv) }$
in Eq.~\eqref{recalc_F1_F2}.
We must evaluate
\begin{equation}
J (\bk , \bv , \bvp) = \!\! \int \!\! \frac{\rd \omega}{2 \pi} \frac{\rd \omegap}{2 \pi} \, \frac{1}{\eps (\bk , \omega) \, \eps (- \bk , \omegap)} \, \frac{\re^{- \ri (\omega + \omegap)t}}{(\omega - \bk\!\cdot\! \bv) (\omega - \bk\!\cdot\!\bvp) (\omegap + \bk\!\cdot\!\bvp)} .
\label{intLaplace_F2}
\end{equation}
Using the same approach as in Eq.~\eqref{intLaplace_F1},
we first perform the integral over $\omegap$, noting that it involves a single pole along the real axis at ${ \omegap = - \bk\!\cdot\!\bvp }$, the other poles being damped. Equation~\eqref{intLaplace_F2} then becomes
\begin{equation}
J (\bk , \bv , \bvp) = - \ri \frac{\re^{\ri \bk \cdot \bvp t}}{\eps^{*} (\bk , \bk \!\cdot\! \bvp)} \! \int \! \frac{\rd \omega}{2 \pi} \, \frac{1}{\eps (\bk , \omega)} \, \frac{\re^{-\ri \omega t}}{(\omega - \bk\!\cdot\! \bv) (\omega - \bk\!\cdot\!\bvp)}
\label{calc_Laplace_F2}
\end{equation}
For the remaining integral there are two poles on the real axis, which  each contribute ${ - 2 \pi \ri }$ times the associated residue. We obtain
\begin{align}
J (\bk , \bv , \bvp) {} & = - \frac{\re^{\ri \bk\cdot\bvp t}}{\eps^{*} (\bk , \bk \!\cdot\! \bvp)} \bigg[ \frac{1}{\eps (\bk , \bk\!\cdot\!\bv)} \, \frac{\re^{- \ri \bk\cdot\bv t}}{\bk\!\cdot\! (\bv - \bvp)} + \frac{1}{\eps (\bk , \bk\!\cdot\! \bvp)} \frac{\re^{- \ri \bk\cdot\bvp t}}{\bk\!\cdot\! (\bvp - \bv)} \bigg]
\nonumber
\\
{} & = \frac{1}{|\eps (\bk , \bk\!\cdot\!\bvp)|^{2}} \, \, \frac{1}{\bk\!\cdot\! (\bv - \bvp)} \bigg[ 1 - \frac{\eps (\bk , \bk\!\cdot\!\bvp)}{\eps (\bk , \bk\!\cdot\! \bv)} \re^{- \ri \bk\cdot (\bv - \bvp) t} \bigg] .
\label{calc_Laplace_F2_2}
\end{align}
Relying on our assumption of timescale separation,
i.e.,\ the assumption that the fluctuations evolve on timescales
much faster than the mean system,
we can take the limit ${ t \to \infty }$ of Eq.~\eqref{calc_Laplace_F2_2}. We then use the identity
\begin{equation}
\lim_{t\to\infty}\frac{\re^{-\ri xt}}{x} = - \ri \pi \deltaD (x) 
\end{equation}
to simplify Eq.~\eqref{calc_Laplace_F2_2} to
\begin{equation}
J (\bk , \bv , \bvp) = \frac{1}{|\eps (\bk , \bk\!\cdot\! \bvp)|^{2}} \, \bigg\{ \mP \bigg[\! \frac{1}{\bk \!\cdot\! (\bv - \bvp)} \!\bigg] + \ri \pi \deltaD [\bk \!\cdot\! (\bv - \bvp)] \bigg\} .
\label{calc_Laplace_F2_3}
\end{equation}

\bigskip

\section{Computing the correlations of the potential fluctuations}
\label{sec:Correlation}

In Section~\ref{sec:DerivationKlim}, we showed that the evolution of the \FDM\ halo \DF\
is sourced by the correlations of the initial fluctuations in the system.
These correlations are described by the function ${ \hC (\bk , \bv) }$, the Fourier transform of the correlation function, as defined in Eq.~\eqref{eq:correl_split}. Let us now explicitly compute this correlation.

In order to introduce that calculation, let us start by considering the classical case.
In that regime, the system's discrete \DF\ is given by
\begin{equation}
F_{\rd} (\br , \bv , t) = \mb \sum_{i = 1}^{N} \deltaD [\br - \br_{i} (t)] \, \deltaD [\bv - \bv_{i} (t)] ,
\label{def_Fd}
\end{equation}
where at the initial time, the phase-space positions and velocities of the particles are drawn independently from another, uniformly in space, and according to the \DF\@ ${ \Fb (\bv) }$ for their velocities.
Similarly to Eq.~\eqref{expansion_W}, the instantaneous fluctuations
in the system's \DF\ are given by ${ f = \Fd - \Fb }$.
At the initial time, we can then write
\begin{equation}
\big\langle f_{0} (\br , \bv) \, f_{0} (\brp , \bvp) \big\rangle = \mb^{2} \sum_{i , j} \big\langle \deltaD (\br \!-\! \br_{i}) \, \deltaD (\bv \!-\! \bv_{i}) \, \deltaD (\brp \!-\! \br_{j}) \, \deltaD (\bvp \!-\! \bv_{j}) \big\rangle - \Fb (\bv) \, \Fb (\bvp) ,
\label{calc_av_classic}
\end{equation}
where we dropped the time dependence (${ t = 0 }$) to shorten the notation.
As the particles are chosen independently, there are two types of terms in the double sum,
depending on whether ${ i = j }$ or ${ i \neq j }$.
We then get
\begin{align}
\big\langle f_{0} (\br , \bv) \, f_{0} (\brp , \bvp) \big\rangle = {} & \mb^{2} \, \deltaD (\br \!-\! \brp) \, \deltaD (\bv \!-\! \bvp) \, \sum_{i} \big\langle \deltaD (\br \!-\! \br_{i}) \, \deltaD (\bv \!-\! \bv_{i}) \big\rangle
\nonumber
\\
{} & + \mb^{2} \sum_{i \neq j} \big\langle \deltaD (\br \!-\! \br_{i}) \, \deltaD (\bv \!-\! \bv_{i}) \big\rangle \, \big\langle \deltaD (\brp \!-\! \br_{j}) \, \deltaD (\bvp \!-\! \bv_{j}) \big\rangle
\nonumber
\\
{} & - \Fb (\bv) \, \Fb (\bvp) .
\label{recalc_av_classic}
\end{align}
Since $\Fb$ obeys the normalisation convention
${ \!\int\! \rd \bv \Fb (\bv) = \rho_{0} }$, we have 
\begin{equation}
\big\langle \deltaD (\br \!-\! \br_{i}) \, \deltaD (\bv \!-\! \bv_{i}) \big\rangle = \frac{1}{N \mb} \, \Fb (\bv) .
\label{prop_avg}
\end{equation}
As a consequence, in the limit ${ N \gg 1 }$, the last two terms in 
Eq.~\eqref{recalc_av_classic} cancel and we have 
\begin{equation}
\big\langle f_{0} (\br , \bv) \, f_{0} (\brp , \bvp) \big\rangle = \mb \Fb (\bv) \, \deltaD (\br \!-\! \brp) \, \deltaD (\bv \!-\! \bvp) . 
\label{rerecalc_av_classic}
\end{equation}
Following the convention from Eq.~\eqref{def_hPhi},
this can be rewritten in Fourier space as
\begin{equation}
\big\langle \hf_{0} (\bk , \bv) \, \hf_{0} (\bkp , \bvp) \big\rangle = \frac{1}{(2 \pi)^{3}} \, \mb \, \Fb (\bv) \, \deltaD (\bk \!+\! \bkp) \, \deltaD (\bv \!-\! \bvp) ,
\label{correl_classic_Fourier}
\end{equation}
and the needed correlation function from Eq.~\eqref{eq:correl_split} is then
\begin{equation}
\hC (\bk , \bv) = \frac{1}{(2 \pi)^{3}} \, \mb \, \Fb (\bv) .
\label{hC_classic}
\end{equation}
We note that this correlation function is independent of $\bk$, a consequence of our assumption that the initial positions of the particles are chosen independently from a homogeneous distribution.

Let us now adapt this calculation to the \FDM\ case and compute the statistics of the persistent fluctuations present in the Wigner distribution function of the halo. We consider the following wavefunction
\begin{equation}
\psi (\br , t) = \!\! \int \!\! \rd \bk \, \vphi (\bk) \, \re^{\ri [\bk \cdot \br - \omega (k) t]} .
\label{psi_hC}
\end{equation}
In the limit where the potential fluctuations ${ \Phi (\br , t) }$ vanish,
this wavefunction is a solution of the free Schr\"{o}dinger equation provided that it satisfies the dispersion relation
\begin{equation}
\omega (k) = \frac{\hbar k^{2}}{2 \mb} .
\label{dispersion_relation}
\end{equation}
Let us now assume that the wavefunction in $\bk$--space,
${ \vphi (\bk) }$, is the sum of Gaussian wavepackets of the form
\begin{equation}
\vphi (\bk) = A \sum_{i = 1}^{N} \re^{\ri \phi_{i}} \, \re^{- \ri \bk \cdot \br_{i}} \, \re^{- |\bk - \mb \bv_{i}/\hbar|^{2} \veps^{2}} .
\label{vphi_hC}
\end{equation}
In that expression, ${ \{ \br_{i} , \bv_{i} \} }$ are random positions and velocities
drawn independently from the \DF\@ ${ \Fb (\bv) }$,
and ${ \{ \phi_{i} \} }$ are independent random phases.
In addition, $\veps$ is an ad hoc parameter,
so that $\veps$ and ${ \hbar / (2 \mb \veps) }$ are respectively
the initial uncertainties in the positions and velocities.
This parameter will prove to be useful in managing our asymptotic developments.
In Eq.~\eqref{vphi_hC}, we also introduced the prefactor $A$
that is tuned to satisfy the normalisation condition stemming from Eq.~\eqref{eq:wigint},
namely that ${ \langle |\psi (\br , t) |^{2} \rangle = \rho_{0} }$.

Let us now determine the value of $A$.
Starting from Eq.~\eqref{psi_hC}, we write
\begin{align}
\big\langle | \psi (\br , t) |^{2} \big\rangle {} & = \!\! \int \!\! \rd \bk \rd \bkp \, \langle \vphi (\bk) \, \vphi^{*} (\bkp) \rangle \, \re^{\ri (\bk - \bkp)\cdot\br } \, \re^{- \ri t [\omega (k) - \omega (\kp)]} .
\label{calc_A}
\end{align}
The two-point correlation function of ${ \vphi (\bk) }$
is 
\begin{align}
\big\langle \vphi (\bk) \, \vphi^{*} (\bkp) \big\rangle {} & = A^{2} \sum_{i , j} \big\langle \re^{\ri (\phi_{i} - \phi_{j})} \, \re^{- \ri \bk \cdot \br_{i}} \, \re^{\ri \bkp \cdot \br_{j}} \, \re^{- |\bk - \mb \bv_{i} / \hbar|^{2} \veps^{2}} \, \re^{- |\bkp - \mb \bv_{j} / \hbar |^{2} \veps^{2}} \big\rangle
\nonumber
\\
{} & = A^{2} \sum_{i} \big\langle \re^{-\ri (\bk - \bkp) \cdot \br_{i}} \, \re^{- |\bk - \mb \bv_{i} / \hbar|^{2} \veps^{2}} \, \re^{- |\bkp - \mb \bv_{i} / \hbar |^{2} \veps^{2}} \big\rangle 
\nonumber
\\
{} & = A^{2} \sum_{i} \!\! \int \!\! \rd \br \rd \bv \, \big\langle \deltaD (\br \!-\! \br_{i}) \, \deltaD (\bv \!-\! \bv_{i}) \big\rangle \, \re^{- \ri (\bk - \bkp) \cdot \br} \, \re^{- |\bk - \mb \bv / \hbar|^{2} \veps^{2}} \, \re^{- |\bkp - \mb \bv / \hbar |^{2} \veps^{2}} .
\label{two_point_vphi}
\end{align}
To get the second line we noted that ${ \langle \re^{\ri (\phi_{i} - \phi_{j})} \rangle }$
is non-zero only for ${ i = j }$.
Using Eq.~\eqref{prop_avg}, one can write
\begin{align}
\big\langle \vphi (\bk) \, \vphi^{*} (\bkp) \big\rangle {} & = \frac{A^{2}}{\mb} \! \int \!\! \rd \br \rd \bv \, \Fb (\bv) \, \re^{- \ri (\bk - \bkp) \cdot \br} \, \re^{- |\bk - \mb \bv / \hbar|^{2} \veps^{2}} \, \re^{- |\bkp - \mb \bv / \hbar |^{2} \veps^{2}}
\nonumber
\\
{} & = \frac{(2 \pi)^{3} \, A^{2}}{\mb} \deltaD (\bk \!-\! \bkp) \!\! \int \!\! \rd \bv \, \Fb (\bv) \, \re^{- 2 |\bk - \mb \bv / \hbar|^{2} \veps^{2}} .
\label{re_two_point_vphi}
\end{align}
We can now use this result to pursue the simplification of Eq.~\eqref{calc_A}. We have
\begin{equation}
\big\langle |\psi (\br , t) |^{2} \big\rangle {}  = \frac{(2 \pi)^{3} A^{2}}{\mb} \!\! \int \!\! \rd \bk \rd \bv \, \Fb (\bv) \, \re^{- 2 |\bk - \mb \bv / \hbar|^{2} \veps^{2}} = \frac{2^{3/2} \pi^{9/2} A^{2}}{\veps^{3} \, \mb} \!\! \int \!\! \rd \bv \, \Fb (\bv) .
\label{re_calc_A}
\end{equation}
Recalling from Eq.~\eqref{eq:wigint} that  ${ \!\int\! \rd \bv \, \Fb(\bv) \!=\! \rho_{0} }$,
we obtain the value of $A$ as
\begin{equation}
A = \frac{\veps^{3/2} \, \mb^{1/2}}{2^{3/4} \pi^{9/4}} .
\label{val_A}
\end{equation}

Relying on the definition from Eq.~\eqref{eq:wigner_I},
we can now compute the Wigner function ${ W (\br , \bv , t) }$
associated with Eq.~\eqref{psi_hC}.
Following some cumbersome calculations,
it takes the form
\begin{align}
W (\br , \bv , t) {} & = \!\! \int \!\! \frac{\rd \bs}{(2 \pi)^{3}} \, \re^{- \ri \bv \cdot \bs} \, \psi \big( \br + \half \hbar \bs/\mb , t \big) \, \psi^{*} (\br - \half\hbar\bs/\mb , t)
\nonumber
\\
{} & = \bigg( \frac{2 \mb}{\hbar} \bigg)^{3} \!\! \int \!\! \rd \bk_{1} \rd \bk_{2} \, \vphi (\bk_{1}) \, \vphi^{*} (\bk_{2}) \, \deltaD \left( \bk_{1} \!+\! \bk_{2} \!-\! \frac{2 \mb}{\hbar} \bv \right) \, \re^{\ri (\bk_{1} - \bk_{2}) \cdot (\br - \bv t)}
\nonumber
\\
{} & = \mb \bigg( \frac{\mb}{\pi \hbar} \bigg)^{3} \sum_{i , j} \re^{\ri (\phi_{i} - \phi_{j})} \, \exp \!\big[\! - \tfrac{2 \veps^{2} \mb^{2}}{\hbar^{2}} | \half(\bv_{i} + \bv_{j}) - \bv |^{2} \big] \, \exp \!\big[\! - \tfrac{1}{2 \veps^{2}} | \half(\br_{i} + \br_{j}) - \br + \bv t |^{2} \big]
\nonumber
\\
{} & \qquad \qquad \qquad \times \exp \!\big\{\! - \ri \tfrac{\mb}{\hbar} \big[ \big( \br_{i} - \br_{j} \big) \!\cdot\! \bv + \big[ \half(\br_{i} + \br_{j}) - \br + \bv t \big] \!\cdot\! ( \bv_{i} - \bv_{j}) \big] \big\} .
\label{wigner_phi}
\end{align}
In the second line we have used the relation $\omega(k_2)-\omega(k_1)=\half(\hbar/\mb)(k_2^2-k_1^2)=\half(\hbar/\mb)(\bk_2+\bk_1)\!\cdot\!(\bk_2-\bk_1)$; when multiplied by the delta function in that equation this reduces to $\bv\cdot(\bk_2-\bk_1)$. 

In order to check these calculations,
let us now follow the same method as in Eq.~\eqref{two_point_vphi}
to compute the ensemble-averaged Wigner function.
Owing to the presence of the factor ${ \re^{\ri (\phi_{i} - \phi_{j})} }$
in Eq.~\eqref{wigner_phi},
only the contributions with ${ i = j }$ remain.
All in all, we obtain
\begin{align}
\oW (\br , \bv , t) {} & = \bigg( \frac{\mb}{\pi \hbar} \bigg)^{3} \!\! \int \!\! \rd \brp \, \rd \bvp \, \Fb (\bvp) \, \exp \!\big[\! - \tfrac{2 \veps^{2} \mb^{2}}{\hbar^{2}} |\bvp - \bv|^{2} \big] \, \exp \!\big[\! - \tfrac{1}{2 \veps^{2}} | \brp - \br + \bv t |^{2} \big]
\nonumber
\\
{} & \simeq \Fb (\bv) ,
\label{calc_av_Wigner_Phi}
\end{align}
To get the last line, we assumed that
${ \veps \gg \lbarsig = \hbar / (\mb \sigma) }$,
with $\sigma$ the system's typical velocity; physically, this means that the size of the wavepacket is much larger than the typical de Broglie wavelength. 
In that limit, we can then use the replacement
\begin{equation}
\re^{- \alpha |\bv|^{2}/\sigma^2} \xrightarrow[\alpha \gg 1]{}  \frac{\pi^{3/2}}{\alpha^{3/2}} \, \deltaD (\bv/\sigma) ,
\label{replacement_Gaussian}
\end{equation}
with ${ \alpha^{-1} \!=\! \hbar^{2}/(2 \veps^{2} \mb^{2} \sigma^{2}) }$
our small parameter. Thus we have recovered in Eq.~\eqref{calc_av_Wigner_Phi} the known result 
that the ensemble-averaged Wigner function
is the same as the \DF\@.

Having computed the Wigner distribution in Eq.~\eqref{wigner_phi},
we can now find the correlation of its fluctuations at the initial time,
as required by Eq.~\eqref{eq:correl_split}.
Following Eq.~\eqref{expansion_W},
we write
\begin{align}
\big\langle f_{0} (\br , \bv) \, f_{0} (\brp , \bvp) \big\rangle {} & = \big\langle W (\br , \bv) \, W (\brp , \bvp) \big\rangle - \oW (\br , \bv) \, \oW (\brp , \bvp) ,
\label{calc_correl_Wigner}
\end{align}
where all the functions are evaluated at the initial time.
Glancing back at Eq.~\eqref{wigner_phi},
we note that the Wigner function can be rewritten in the shorter form
\begin{equation}
W (\br , \bv ) = \sum_{i , j} \re^{\ri (\phi_{i} - \phi_{j})} \, g_{ij} (\br , \bv) ,
\label{shape_Wigner}
\end{equation}
where the expression for the function ${ g_{ij} (\br , \bv) }$
naturally follows from Eq.~\eqref{wigner_phi}.
As a result, the ensemble average in the r.h.s.\ of Eq.~\eqref{calc_correl_Wigner}
takes the form
\begin{equation}
\big\langle W (\br , \bv) \, W (\brp , \bvp) \big\rangle = \sum_{\substack{i,j\\k,l}} \big\langle \re^{\ri (\phi_{i} - \phi_{j} + \phi_{k} - \phi_{l})} \, g_{ij} (\br , \bv) \, g_{kl} (\brp , \bvp) \big\rangle
\label{shape_average}
\end{equation}
Because the wavepackets are drawn independently, the phase term ${ \langle \re^{\ri (\phi_{i} - \phi_{j} + \phi_{k} - \phi_{l})} \rangle }$ is non-zero only in three cases, namely
(i) ${ i \!=\! j \!=\! k \!=\! l }$;
(ii) ${ i \!=\! j }$ and ${ k \!=\! l }$ with ${ i \!\neq\! k }$;
(iii) ${ i \!=\! l }$ and ${ j \!=\! k }$ with ${ i \!\neq\! k }$.
In the limit ${ N \gg 1 }$,
we can then rewrite Eq.~\eqref{shape_average} as
\begin{equation}
\big\langle W (\br , \bv) \, W (\brp , \bvp) \big\rangle = N \, \big\langle g_{ii} (\br , \bv) \, g_{ii} (\brp , \bvp) \big\rangle 
+ N^{2} \big\langle g_{ii} (\br , \bv) \, g_{kk} (\brp , \bvp) \big\rangle
+  N^{2} \big\langle g_{ik} (\br , \bv) \, g_{ki} (\brp , \bvp) \big\rangle ,
\label{rewrite_shape_average}
\end{equation}
where, in the last two terms,
it is understood that ${ \{ \br_{i} , \bv_{i} \} }$
and ${ \{ \br_{k} , \bv_{k} \} }$ are two independent sets of random variables.
Let us now compute in turn each of the terms appearing in Eq.~\eqref{rewrite_shape_average}.
We can first write
\begin{align}
N \big\langle g_{ii} (\br , \bv) \, g_{ii} (\brp , \bvp) \big\rangle = {} & \mb
 \bigg( \frac{\mb}{\pi \hbar} \bigg)^{6} \!\! \int \!\! \rd \bv_{i} \, \Fb (\bv_{i}) \, \exp \!\big[\! - \tfrac{2 \veps^{2} \mb^{2}}{\hbar^{2}} \big( |\bv_{i} - \bv|^{2} + |\bv_{i} - \bvp|^{2} ) \big]
 \nonumber
 \\
{} & \qquad \qquad \times \!\! \int \!\! \rd \br_{i} \exp \!\big[\! - \tfrac{1}{2 \veps^{2}} \big( |\br_{i} - \br|^{2} + |\br_{i} - \brp|^{2} \big) \big]
\nonumber
\\
= {} & \mb \bigg( \frac{\mb}{\pi \hbar} \bigg)^{6} \exp \!\big[\! - \tfrac{\veps^{2} \mb^{2}}{\hbar^{2}} |\bv - \bvp|^{2} \big] \!\!\int\!\! \rd \bv_{i} \, \Fb (\bv_{i}) \, \exp \!\big[\! - \tfrac{4 \veps^{2} \mb^{2}}{\hbar^{2}} |\bv_{i} - \half(\bv + \bvp) |^{2} \big]
\nonumber
\\
{} & \qquad \qquad\times \exp \!\big[\! - \tfrac{1}{4 \veps^{2}} | \br - \brp |^{2} \big] \!\! \int \!\! \rd \br_{i} \, \exp \!\big[\! - \tfrac{1}{\veps^{2}} |\br_{i} - \half(\br + \brp)|^{2} \big]
\nonumber
\\
\simeq {} & \mb \, \deltaD (\br \!-\! \brp) \, \deltaD (\bv \!-\! \bvp) \, \oW (\bv) ,
 \label{av_1}
\end{align}
where to get the last line,
we assumed once again that ${\veps \gg \lbarsig }$,
and used the asymptotic replacement from Eq.~\eqref{replacement_Gaussian}.
The second term from Eq.~\eqref{rewrite_shape_average} reads
\begin{align}
N^{2} \big\langle g_{ii} (\br , \bv) \, g_{kk} (\brp , \bvp) \big\rangle = {} & \bigg( \frac{\mb}{\pi \hbar} \bigg)^{3} \!\! \int \!\! \rd \br_{i} \rd \bv_{i} \, \Fb (\bv_{i}) \, \exp \!\big[\! - \tfrac{2 \veps^{2} \mb^{2}}{\hbar^{2}} |\bv_{i} - \bv|^{2} \big] \, \exp \!\big[\! - \tfrac{1}{2 \veps^{2}} |\br_{i} - \br|^{2} \big]
\nonumber
\\
{} & \qquad\times\bigg( \frac{\mb}{\pi \hbar} \bigg)^{3} \!\! \int \!\! \rd \br_{k} \rd \bv_{k} \, \Fb (\bv_{k}) \, \exp \!\big[\! - \tfrac{2 \veps^{2} \mb^{2}}{\hbar^{2}} | \bv_{k} - \bvp |^{2} \big] \, \exp \!\big[\! - \tfrac{1}{2 \veps^{2}} |\br_{k} - \brp|^{2} \big]
\nonumber
\\
= {} & \oW (\br , \bv) \, \oW (\brp , \bvp) ,
\label{av_2}
\end{align}
where we used the result from Eq.~\eqref{calc_av_Wigner_Phi}.
The last term from Eq.~\eqref{rewrite_shape_average} then reads
\begin{align}
N^{2} \big\langle g_{ik} (\br, \bv) \, g_{ki} (\brp , \bvp) \big\rangle = {} & \bigg( \frac{\mb}{\pi \hbar} \bigg)^{6} \!\! \int \!\! \rd \br_{i} \rd \bv_{i} \rd \br_{k} \rd \bv_{k} \, \Fb (\bv_{i}) \, \Fb (\bv_{k}) \, \exp \!\big[\! - \tfrac{2 \veps^{2} \mb^{2}}{\hbar^{2}} \big( | \half(\bv_{i} + \bv_{k}) - \bv |^{2} + | \half(\bv_{i} + \bv_{k}) - \bvp |^{2} \big) \big]
\nonumber
\\
{} & \qquad \qquad \times \exp \!\big[\! - \tfrac{1}{2 \veps^{2}} \big( | \half(\br_{i} + \br_{k}) - \br |^{2} + | \half(\br_{i} + \br_{k}) - \brp |^{2}  \big) \big]
\nonumber
\\
{} & \qquad \qquad \times \exp \!\big\{\! - \ri \tfrac{\mb}{\hbar} \big[ ( \br_{i} \!-\! \br_{k} ) \!\cdot\! ( \bv \!-\! \bvp ) \!-\! (\br \!-\! \brp) \!\cdot\! (\bv_{i} \!-\! \bv_{k}) \big] \big\}
\nonumber
\\
= {} & \bigg( \frac{\mb}{\pi \hbar} \bigg)^{6} \, \exp \!\big[\! - \tfrac{\veps^{2} \mb^{2}}{\hbar^{2}} | \bv - \bvp |^{2} \big] \, \exp \!\big[\! - \tfrac{1}{4 \veps^{2}} |\br - \brp |^{2} \big]
\nonumber
\\
{} & \quad \times \!\! \int \!\! \rd \br_{i} \rd \bv_{i} \rd \br_{k} \rd \bv_{k} \, \Fb (\bv_{i}) \, \Fb (\bv_{k}) \, \exp \!\big[\! - \tfrac{4 \veps^{2} \mb^{2}}{\hbar^{2}} | \half(\bv_{i} + \bv_{k}) - \half(\bv + \bvp) |^{2} \big]
\nonumber
\\
{} & \quad \times \exp \!\big[\! - \tfrac{1}{\veps^{2}} | \half(\br_{i} + \br_{k}) - \half(\br + \brp) |^{2} \big] \, \exp \!\big\{\! -\! \ri \tfrac{\mb}{\hbar} \big[ (\br_{i} \!-\! \br_{k}) \!\cdot\! (\bv \!-\! \bvp) \!-\! (\br \!-\! \brp) \!\cdot\! (\bv_{i} \!-\! \bv_{k}) \big] \big\} .
\label{av_3}
\end{align}
At this stage, the asymptotic formula from Eq.~\eqref{replacement_Gaussian}
has to be used carefully,
because the complex exponential from Eq.~\eqref{av_3}
is rapidly fluctuating.
To clarify this calculation, let us perform the change of variables
\begin{equation}
\begin{cases}
\displaystyle \bsigma_{\br} = \half(\br_{i} + \br_{k}),
\\
\displaystyle \bdelta_{\br} = \br_{i} - \br_{k} ,
\end{cases}
; \quad
\begin{cases}
\displaystyle \bsigma_{\bv} = \half(\bv_{i} + \bv_{k}) ,
\\
\displaystyle \bdelta_{\bv} = \bv_{i} - \bv_{k} .
\end{cases}
\label{change_var_av_3}
\end{equation}
Equation~\eqref{av_3} becomes
\begin{align}
N^{2} \big\langle g_{ik} (\br, \bv) \, g_{ki} (\brp , \bvp) \big\rangle  & = \bigg( \frac{\mb}{\pi \hbar} \bigg)^{6} \exp \!\big[\! - \tfrac{\veps^{2} \mb^{2}}{\hbar} |\bv - \bvp|^{2} \big] \, \exp \!\big[\! - \tfrac{1}{4 \veps^{2}} |\br - \brp|^{2} \big]
\nonumber
\\
{} & \times \!\! \int \!\! \rd \bsigma_{\bv} \rd \bdelta_{\bv} \, \Fb (\bsigma_{\bv} \!+\! \half \bdelta_{\bv}) \, \Fb (\bsigma_{\bv} \!-\! \half \bdelta_{\bv}) \, \exp \!\big[\! - \tfrac{4 \veps^{2} \mb^{2}}{\hbar^{2}} | \bsigma_{\bv} \!-\! \half(\bv \!+\! \bvp) |^{2} \big] \, \exp \!\big[\! - \ri \tfrac{\mb}{\hbar} (\br \!-\! \brp) \!\cdot\! \bdelta_{\bv} \big]
\nonumber
\\
{} & \times \!\! \int \!\! \rd \bsigma_{\br} \rd \bdelta_{\br} \, \exp \!\big[\! - \tfrac{1}{\veps^{2}} | \bsigma_{\br} - \half(\br + \brp) |^{2} \big] \, \exp \!\big[\! - \ri \tfrac{\mb}{\hbar} \bdelta_{\br} \!\cdot\! (\bv - \bvp) \big] .
\label{re_av_3}
\end{align}
In this expression, we note that the exponential factor ${ \exp [- \tfrac{1}{4 \veps^{2}} |\br \!-\brp|^{2}] }$ is nearly zero unless
${ |\br \!-\! \brp | \lesssim \veps }$.
In that regime, the complex exponential ${ \exp [ - \ri \tfrac{\mb}{\hbar} (\br \!-\! \brp) \!\cdot\! \bdelta_{\bv} ] }$
will average to nearly zero unless 
${ |\bdelta_{\bv}| \lesssim \hbar/(\veps \mb) }=\sigma\lbarsig/\epsilon$.
Given our assumption that ${ \veps \gg \lbarsig }$, we conclude that the dominant contribution to the integral comes from ${ |\bdelta_{\bv}| \ll \sigma }$.
We recall that the typical variance of ${ \Fb (\bv) }$ is $\sigma$, so in Eq.~\eqref{re_av_3}
we may perform the replacement
${ \Fb (\bsigma_{\bv} \!+\! \half \bdelta_{\bv}) \Fb (\bsigma_{\bv} \!-\! \half \bdelta_{\bv}) \!\to\! \Fb^{2} (\bsigma_{\bv}) }$.
We get
\begin{align}
N^{2} \big\langle g_{ik} (\br, \bv) \, g_{ki} (\brp , \bvp) \big\rangle \simeq {} & 2^{6} \, \deltaD (\br \!-\! \brp) \, \deltaD (\bv \!-\! \bvp) \!\! \int \!\! \rd \bsigma_{\br} \rd \bsigma_{\bv} \, \exp \!\big[\! - \tfrac{1}{\veps^{2}} |\bsigma_{\br} - \br|^{2} \big] \, \Fb^{2} (\bsigma_{\bv}) \, \exp \!\big[\! - \tfrac{4 \veps^{2} \mb^{2}}{\hbar^{2}} |\bsigma_{\bv} - \bv|^{2} \big]
\nonumber
\\
\simeq {} & \frac{h^{3}}{\mb^{3}} \, \deltaD (\br \!-\! \brp) \, \deltaD (\bv \!-\! \bvp) \, \oW^{2} (\bv) ,
\label{re_av_3bis}
\end{align}
Gathering together Eqs.~\eqref{av_1},~\eqref{av_2},
and~\eqref{re_av_3},
we can now rewrite the correlation from Eq.~\eqref{calc_correl_Wigner}
to obtain
\begin{equation}
\big\langle f_{0} (\br , \bv) \, f_{0} (\brp , \bvp) \big\rangle = \bigg[ \mb + \frac{h^{3}}{\mb^{3}} \, \oW (\bv) \bigg] \, \oW (\bv) \, \deltaD (\br \!-\! \brp) \, \deltaD (\bv \!-\! \bvp) .
\label{calc_correl_Wigner_final}
\end{equation}
Following Eqs.~\eqref{def_hPhi} and~\eqref{eq:correl_split},
we finally obtain the needed correlation function as
\begin{equation}
\hC (\bk , \bv) = \frac{1}{(2 \pi)^{3}} \bigg[ \mb + \frac{h^{3}}{\mb^{3}} \oW (\bv) \bigg] \, \oW (\bv) ,
\label{hC_fuzzy}
\end{equation}
which reduces to the classical correlation function \eqref{hC_classic} as $h\to 0$.

\end{document}